\documentclass[twocolumn,showpacs,prd]{revtex4}
\usepackage{graphicx} 
\usepackage{amsmath} 
\usepackage{amssymb}
\usepackage{mathrsfs}
\usepackage{psfrag}
\newtheorem{conjecture}{Conjecture}

\newcommand{\dv}{\, d\mu}

\newcommand{\Rt}{\mathbb{R}^3}

\newcommand{\Rd}{\mathbb{R}^2}

\newcommand{\mf}{\mathcal{M}}

\newcommand{\X}{\eta}

\newcommand{\Y}{\omega}

\newcommand{\x}{\sigma}

\begin{document}

\title{Numerical evidences for the  angular momentum-mass inequality
  for multiple axially symmetric black holes}

\author{Sergio Dain}
\affiliation{Facultad de Matem\'atica,
  Astronom\'{\i}a y F\'{i}sica, Universidad Nacional de C\'ordoba,
  Ciudad Universitaria (5000) C\'ordoba, Argentina;}
\affiliation{Max Planck Institute for Gravitational Physics (Albert
  Einstein Institute) Am M\"uhlenberg 1 D-14476 Potsdam Germany.}

\author{Omar Ortiz}
\affiliation{Facultad de Matem\'atica,
  Astronom\'{\i}a y F\'{i}sica, Universidad Nacional de C\'ordoba,
  Ciudad Universitaria (5000) C\'ordoba, Argentina.  }

\date{\today}

\begin{abstract}
  We present numerical evidences for the validity of the inequality
  between the total mass and the total angular momentum for multiple
  axially symmetric (non-stationary) black holes. We use a parabolic
  heat flow to solve numerically the stationary axially symmetric
  Einstein equations. As a by product our method, we also give
  numerical evidences that there are no regular solutions of Einstein equations
  that describe two extreme, axially symmetric black holes in
  equilibrium.
\end{abstract}


\maketitle

\section{Introduction}\label{sec:introduction}

The final state of a gravitational collapse is expected to be
described by a black hole and not by a naked singularity. Moreover, at
late times, the system should settle down to a stationary regime and
since the Kerr black hole is expected to be the only stationary black
hole in vacuum, the final state of all possible gravitational
collapses should approach to a Kerr black hole.  For simplicity, in
this discussion we are not considering electromagnetic fields and also
we are assuming that at some finite time all the matter fields have
fallen into the black hole.

The above considerations roughly encompass what is known as the
standard picture of the gravitational collapse which, in particular,
includes the weak cosmic censorship conjecture.  To prove that this
heuristic picture is in fact a consequence of the Einstein field
equations is one of the most relevant open problems in classical
General Relativity.


One fruitful strategy to study some aspects of this problem is the
following.  From the heuristic picture presented above it is possible
to deduce some geometric inequalities on the initial conditions for
gravitational collapse.  Hence, initial conditions that violate these
inequalities would automatically provide counter examples for the
validity of the standard picture of the gravitational collapse. In
fact, the original intention of this strategy, proposed first by
Penrose \cite{Penrose73}, was to construct such counter examples.
However it was not possible to find them. It
was then natural trying to prove these inequalities. Such proofs provide an
indirect but highly non trivial evidence that the heuristic picture of
the gravitational collapse is correct (see the discussion in
\cite{Wald99}).  This kind of inequalities are also interesting by
themselves because they provide unexpected mathematical connections
between geometric quantities.

A prominent example of this idea is the Penrose inequality which
relates the mass with the area of the black hole horizon on the
initial conditions. An important special case of this inequality has
been proved in \cite{Huisken01} \cite{Bray01} (see also the review
article \cite{Bray04}).  Another example of this kind of inequalities
is the inequality between mass and angular momentum. This inequality,
which constitute the main subject of the present article, arises as
follows.


Consider an axially symmetric gravitational collapse.  An important
feature of axial symmetry is that axially symmetric waves can not
carry angular momentum.  In other words: in vacuum, angular momentum
is a conserved quantity in axial symmetry. Let us assume that the
heuristic picture presented above is correct.  Denote by $m_0$ and
$J_0$ the mass and angular momentum of the final Kerr black hole. The
Kerr black hole satisfies the inequality
\begin{equation} 
\label{eq:inqKerr} 
\sqrt{|J_0|}\leq m_0.  
\end{equation} 
The Kerr solution is well defined for any choice of the parameters $m_0$ and
$J_0$, it defines however a black hole only if inequality \eqref{eq:inqKerr} is
satisfied. Let $m$ and $J$ be the total mass and total angular momentum of the
initial conditions. Since gravitational waves carry positive mass we have
$m_0\leq m$ (this inequality is of course also valid without the assumption of
axial symmetry).  And because angular momentum is conserved in axial symmetry we
have $J=J_0$. Hence, in order to reach the inequality \eqref{eq:inqKerr} at a late
time, every initial condition for axially symmetric collapse must satisfy
\begin{equation} 
\label{eq:14a} 
\sqrt{|J|}\leq m. 
\end{equation} 
See \cite{Dain05e} for a more detailed physical discussion.  This
inequality involves only quantities defined on the initial
conditions. It is expected to hold for every axially symmetric vacuum
(not necessarily stationary) black hole.  The inequality
\eqref{eq:14a} was studied in a series of articles \cite{Dain05c}
\cite{Dain05d} \cite{Dain05e} and finally proved for the case of one
black hole in \cite{Dain06c} and \cite{Chrusciel:2007ak}. There exists
however no proof for the case of multiple axially symmetric black
holes. This problem appears to be deeper (and considerable more
difficult) than the single black hole case. In particular, it is
related, as we discuss below, with the still open problem of the
uniqueness of the Kerr black hole among stationary black holes with
disconnected horizons.  The main purpose of this article is to provide
numerical evidence for the validity of \eqref{eq:14a} for multiple
black holes.

A naive method to test \eqref{eq:14a} is to take some configuration of
axially symmetric black holes and compute numerically the mass and the
angular momentum of it. For a given configuration the relevant
parameters are the separation distance between the black holes and the
individual angular momentum of them. But, of course, these parameters
(or any other finite set of parameters) do not characterize uniquely
the initial conditions.  There exists infinitely many configurations
with the same parameters, this essentially corresponds to the freedom
of including gravitational waves surrounding the black holes. Then,
either we find a counter example or this naive method will give a very
poor evidence in favor of \eqref{eq:14a}. Just some isolated points in
the space of all possible initial conditions. 

Fortunately a different approach is possible. It is based on the
variational principle for the inequality \eqref{eq:14a} presented in
\cite{Dain05c}. This variational principle states that the minimum of
the mass of a given configuration with fixed angular momentum is
achieved by the associated (i.e. with the same parameters) stationary
and axially symmetric solution of Einstein equations.  Hence, in order
to prove the inequality \eqref{eq:14a} for a given configuration it is
enough to compute the mass of the corresponding stationary and axially
symmetric solution of Einstein equations, which is characterized by the
separation distance and individual angular momentum of the black
holes. The stationary and axially symmetric Einstein equations are non
linear elliptic equations. In this article, we use a heat flow to
numerically solve them.  This parabolic flow has two important
properties, first for arbitrary data it converges (as time goes to
infinity) to a stationary and axially symmetric solutions of Einstein
equations. Second, the mass is monotonically decreasing along the
evolution and the angular momentum is conserved (under appropriate
boundary conditions). Hence, the flow provides an accurate procedure
for computing the minimum of the mass of each possible configuration.
This method is interesting by itself as a method for solving
numerically the stationary axially symmetric Einstein equations with
prescribed boundary conditions, which, up to the best of our
knowledge, have not been used so far.
 
For simplicity we will restrict ourselves to configurations with only two black
holes, although our method applies for any number of black holes. For this
configuration, the most favorable case to violate the inequality \eqref{eq:14a}
is when the black holes have the same angular momentum pointing in the same
direction. This corresponds to a repulsive spin-spin force between them. This is
also the most favorable case for reaching a stationary solution representing two
black holes at equilibrium, because it is in principle conceivable that the
repulsive spin force balance the gravitational attraction. This configuration
has only two parameters, the separation distance and the angular momentum.
However, as we will see in the next section, due to the scale invariance of the
equations we have only one non trivial parameter, which we chose to be the
separation distance. We can compute the mass for every choice of the separation
distance and plot a curve.  From the shape of the curve it is clear that,
although we can compute only a finite range, the inequality will be satisfied for
every separation distance. For other configurations we proceed in similar way.
Then, we obtain fairly strong numerical evidences that the inequality is
satisfied for two black holes with any separation distances and any angular
momentum.

As we mention above, the heat flow relaxes to a solution of the
stationary and axially symmetric Einstein vacuum equations.  An
important open problem in General Relativity is whether the Kerr black
hole is unique among stationary black holes (see the recent article
\cite{Chrusciel:2008js} and reference therein). This is essentially
the same problem as whether is possible to achieve an equilibrium
configuration of multiple black holes in General Relativity.  This
problem have been studied in \cite{Weinstein96}\cite{Weinstein94}
\cite{Weinstein92}\cite{Weinstein90}\cite{Li92} \cite{Li91b}. For some
limit cases and also for cases with reflection symmetry it has been
proved that equilibrium is not possible.  Also, from a different
perspective, the problem has been studied using exact solutions in
\cite{Manko01}. Again, the conclusion was that equilibrium is not
possible for this class of solutions.  Using the heat flow, in this
article we also provide numerical evidences that there is no regular
equilibrium solution for two extreme black holes.  This case has not
been analyzed previously in the literature.

The plan of the article is the following. In section
\ref{sec:vari-probl-parab} we introduce the heat flow and analyze its
main properties. We also discuss the precise form of the conjecture
regarding inequality \eqref{eq:14a} and its relation with the black
hole equilibrium problem mentioned above. In section
\ref{sec:Techniques} we discuss the numerical techniques used to solve
the parabolic heat flow equations. In section \ref{sec:results} we
present our results and in section \ref{sec:conclusion} we give some
further perspective on the open problems.  Finally, for the sake of
completeness, we include an appendix \ref{sec:extr-kerr-solut} with
the explicit form of the extreme Kerr solution used in our
computations.

\section{The variational problem and the parabolic flow}
\label{sec:vari-probl-parab}

Consider  a vacuum, axially symmetric spacetime. The axial Killing vector
defines two geometrical scalars, the square of its norm $\eta$ and the twist
potential $\omega$.  These scalars characterize the spacetime in the following
sense. Take a foliation of Cauchy surfaces on the spacetime with the
corresponding time function. An initial data set for the spacetime is
determined by the value of the functions $(\eta,\omega)$ and the time
derivatives $(\eta',\omega')$ on a Cauchy surface.  The Einstein evolution
equations essentially reduce to a non-linear system of
wave equations for $(\eta,\omega)$.  In appropriate coordinates, the
total mass $m$ of the spacetime can be written as a positive integral on a
Cauchy surface in terms of $(\eta,\omega)$ and $(\eta',\omega')$. This integral
is the non linear and conserved energy of the system of waves equations (see
\cite{Dain:2008xr} for details).  

An initial data set is called ``momentary stationary'' if
$(\eta',\omega')$ vanished. Stationary data are a particular class of
momentary stationary data for which the scalars $(\eta, \omega)$
satisfy a set of elliptic equations (see equations
(\ref{eq:haf1})--(\ref{eq:haf2}) below).  An important feature of the
mass integral is that, for arbitrary data, the associated momentary
stationary data has less or equal mass. That is, there exists a lower
bound for the mass that can be written in terms only on
$(\eta,\omega)$ and no time derivatives $(\eta',\omega')$ are
involved.  This lower bound for the mass plays a key role in order to
reduce the proof of the inequality (\ref{eq:14a}) to a pure
variational problem. It can be written as an integral in $\Rt$ as
follows (for details see \cite{Dain06c}, \cite{Dain:2007pk},
\cite{Dain05c}, \cite{Dain05e}).

\begin{figure}
\begin{center}
\includegraphics[width=4cm]{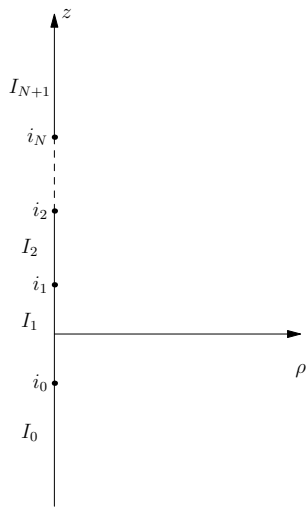}
\caption{$N$ asymptotic ends\label{fig:1}}
\end{center}
\end{figure}

Let $x^i$ be Cartesian coordinates in $\Rt$ (denoted also by $x=x^1$,
$y=x^2$, $z=x^3$) and let $(\rho, \phi)$ be the associated cylindrical
coordinates defined by $\rho=\sqrt{x^2+y^2}$, $\tan\phi = y/x$. The
positions of the black holes will be prescribed by a finite collection
of points $i_k$ located at the $z$ axis. More precisely, the points
$i_k$ will represent extra asymptotic ends on the spacetime and they
can be associated with the location of the black holes. For a given
set of $N+1$ points $i_k$ we define the separation intervals $I_k$,
$0\leq k\leq N-1 $, to be the open sets in the axis between $i_k$ and
$i_{k-1}$, and we define $I_0$ and $I_N$ as $z< i_0$ and $z >i_N$
respectively. Let $L_k$ be the length of $I_k$ for $0\leq k\leq N-1 $.
See figure \ref{fig:1}.  The length $L_k$ (which are measured with
respect to the Cartesian coordinates introduced above) will be
associated with the separation distance between the black holes (see
the discussion in \cite{Dain02e}).

From the square norm $\eta$ of the Killing vector we define the
following function $\sigma$
\begin{equation}
  \label{eq:5}
  \X=\rho^2e^{\x}.
\end{equation}
The lower bound of the mass is given by the following functional
\begin{equation}
  \label{eq:4}
  \mf(\x,\Y)= \frac{1}{32\pi}\int_{\Rt}
  \left(|\partial  \x |^2  +\rho^{-4} e^{-2\x} |\partial \Y |^2
  \right) \dv,
\end{equation}
where $\dv$ is the volume element of $\Rt$, $\partial_i$ denotes
partial derivatives with respect to Cartesian coordinates $x^i$ and
$|\partial \x |^2=\partial_i \x\partial^i \x$. As we mentioned above,
for an arbitrary axially symmetric initial data $(\eta, \omega, \eta',
\omega')$ with mass $m$ we have that (see \cite{Dain06c})
\begin{equation}
  \label{eq:2}
  m\geq \mf.
\end{equation}

The angular momentum $J_k$ of the end $i_k$ is given by
\begin{equation}
  \label{eq:27}
  J_k =\frac{1}{8}\left ( \omega|_{I_{k+1}} - \omega|_{I_{k}}\right ).
\end{equation}  
The total angular momentum is defined by
\begin{equation}
  \label{eq:54}
J=\sum_{k=0}^{N-1} J_k= \frac{1}{8}\left ( \omega|_{I_N} -
\omega|_{I_{0}}\right ).
\end{equation}  
Note the value of the function $\omega$ at the axis prescribe the
angular momentum of the configuration. 

The Euler-Lagrange equations  of the functional $\mf$ are given by 
\begin{align}
  \label{eq:el1}
  \Delta \x   -\frac{e^{-2\x}|\partial \Y |^2}{\rho ^4}& =0,\\
\label{eq:el2}
\partial_i \left(\frac{\partial^i\Y}{\X^2}\right)&=0 , 
\end{align}
In these equations $\Delta=\partial_i\partial^i$ denotes the flat
Laplacian in $\Rt$.  An important property of the functional
\eqref{eq:4} is that equations (\ref{eq:el1})--(\ref{eq:el2})
correspond to the stationary axially symmetric Einstein equations.

We are now in position to formulate the variational approach of the
inequality \eqref{eq:14a}. The conjecture is the following:
\begin{conjecture}
\label{c:1}
 For arbitrary functions $(\sigma, \omega)$ we have
  \begin{equation}
\label{eq:inm}
    \mf (\x,\Y)\geq \sqrt{|J|}, 
  \end{equation}
  where $J$ is given by (\ref{eq:54}).  Moreover, the equality in
  \eqref{eq:inm} implies that the functions $(\x,\Y)$ correspond to
  the extreme Kerr solution. That is, for fixed total angular momentum
  $J$, the extreme Kerr solution is the unique absolute minimum of
  $\mf$.
\end{conjecture}
The inequality \eqref{eq:14a} is a direct consequence of this
conjecture and \eqref{eq:2}.  It is important to emphasize that the
number of end points $i_k$ and their corresponding angular momentum
$J_k$ are not fixed. That is, the conjecture states that for fixed
$J$, extreme Kerr is the unique absolute minimum among all possible
functions $(\sigma, \omega)$ and among all possible configurations of
ends $i_k$ with individual angular momentum $J_k$. Note that in order
to have a non zero $J$ we need at least one end point.

This is a singular variational problem since a non zero $J$ implies
(by equation \eqref{eq:54}) that at least one $J_k$ is non zero, then
equation \eqref{eq:27} implies that $\omega$ is discontinuous at $i_k$
and hence has infinity gradient at this point. In order to make the
second term in the integral \eqref{eq:4} finite the function $\sigma$
should diverge at $i_k$ to compensate the divergence of the gradient
of $\omega$. Also, the singularity of $\sigma$ at $i_k$ can not be too
severe because the first term in the integral \eqref{eq:4} should
remain bounded.

In the formulation of the conjecture we did not specify the functional
space of admissible functions $(\sigma, \omega)$ for the variational
problem. As we mentioned above, the functions are typically singular
at $i_k$, and hence the prescription of the appropriate functional
space can be quite subtle.  We will no discuss this issue here since
it is beyond the scope of this article. For our present purpose, it is
enough to assume some space of admissible functions which is regular
enough in order that the integral \eqref{eq:4} is well defined but it
is also compatible with the singular boundary conditions (\ref{eq:54})
(for a discussion regarding this point see \cite{Dain06c})

Conjecture  \ref{c:1} was proved for the case $N=1$ in \cite{Dain06c} and
\cite{Chrusciel:2007ak}.  The case $N\geq 2$ is open.  Remarkably, for
general $N$ in \cite{Chrusciel:2007ak} it has been proved that if the
ends $i_k$ and the individual angular momentum $J_k$ are fixed then
there exists a unique minimum of the functional $\mf$. This minimum
satisfies the Euler-Lagrange equations (\ref{eq:el1})--(\ref{eq:el2}).
That is, for fixed $i_k$ and $J_k$, there exists functions
$(\sigma_{min},\omega_{min})$, solutions of
(\ref{eq:el1})--(\ref{eq:el2}), where $\omega_{min}$ satisfies
(\ref{eq:27}), such that
\begin{equation}
\label{eq:min0}
\mf(\sigma,\omega) \geq \mf_{min}, 
\end{equation}
for all admissible functions $(\sigma, \omega)$ where $\omega$
satisfies the boundary condition \eqref{eq:27} and we have defined
\begin{equation}
  \label{eq:19}
  \mf_{min} \equiv \mf(\sigma_{min},\omega_{min}). 
\end{equation}
What is not known is
the value of $\mf_{min}$. In particular, it is
not known if this minimum satisfies the inequality (\ref{eq:inm}) for
$N\geq 2$.  A natural strategy to prove the conjecture is to prove
that for arbitrary $i_k$ and $J_k$ the minimum
$\mf_{min}$ satisfies (\ref{eq:inm}).  The main
goal of this article is to compute numerically this value for
different configurations, showing that it satisfies the inequality
\eqref{eq:inm} in all considered cases.

In order to compute $\mf_{min}$ we need to calculate the
solution $(\sigma_{min}, \omega_{min})$ of the Euler-Lagrange equations
(\ref{eq:el1})--(\ref{eq:el2}) with boundary conditions
\eqref{eq:27}. As an efficient method for computing numerically both
the solution and the value of the energy $\mf$ we propose a heat flow
defined as follows.  We consider functions $(\x,\Y)$ which depend on the
coordinates $x^i$ and an extra parameter $t$.  Then, we define the
following flow
\begin{align}
  \label{eq:haf1}
\dot \x  &=  \Delta \x  +\frac{e^{-2\x}|\partial \Y |^2}{\rho ^4},\\
\label{eq:haf2}
\dot \Y  & =\eta^2 \partial_i \left(\frac{\partial^i\Y}{\X^2}\right),   
\end{align}
where a dot denotes partial derivative with respect to $t$.  That is,
we have added time derivatives to the right hand side of equations
(\ref{eq:el1})--(\ref{eq:el2}).  Equations
(\ref{eq:haf1})--(\ref{eq:haf2}) represent the gradient flow of the
energy (\ref{eq:4}).  As $t\to \infty$ we expect that the solution of
the flow will reach a stationary regime (i.e. $\dot\sigma=\dot \Y=0)$
and hence it will provide a solution of equations
(\ref{eq:el1})--(\ref{eq:el2}).

The important property of the flow is that the energy $\mf$ is
monotonic under appropriate boundary conditions. This can be seen as
follows.  Consider the functional (\ref{eq:4}) defined on a bounded
domain $\Omega$ (denoted in the following by $\mf_\Omega$) for
functions that are solutions of (\ref{eq:haf1})--(\ref{eq:haf2}) and
take a time derivative of $\mf_\Omega$. Integrating by parts and using
the evolution equations (\ref{eq:haf1})--(\ref{eq:haf2}) we obtain
\begin{multline}
  \label{eq:10}
  \dot \mf_\Omega= -\frac{1}{16\pi}\int_{\Omega}\left( \dot \x^2 +
   \eta^{-2}  \dot\Y^2 \right)\dv +\\
\frac{1}{16\pi} \oint_{\partial\Omega}\left( \dot \x \partial_n
      \x + \eta^{-2} \dot \Y \partial_n \Y \right) \, ds,
\end{multline}
where $ds$ is the area element of the boundary $\partial \Omega$
and $\partial_n$ denotes exterior normal derivative with
respect to $\partial \Omega$.   
By combining of homogeneous Neumann boundary conditions
\begin{equation}
  \label{eq:42}
 \partial_n \x  =0, \quad \partial_n \Y  =0  \text{ on } \partial \Omega,
\end{equation}
or Dirichlet boundary conditions
\begin{equation}
  \label{eq:11}
  \sigma = g_1, \quad \omega=g_2  \text{ on } \partial \Omega,
\end{equation}
for arbitrary time independent functions $g_1, g_2$ (since in this case
we get $ \dot \sigma = \dot \omega=0$ on the boundary) will make the boundary
term in (\ref{eq:10}) vanish.
And hence we get that 
\begin{equation}
  \label{eq:43}
  \dot \mf_\Omega= -\frac{1}{16\pi}\int_{\Omega}\left( \dot \x^2 +
    \eta^{-2} \dot\Y^2 
  \right)\dv \leq 0. 
\end{equation}
When the domain $\Omega$ is $\Rt$  we need to prescribe
appropriate fall off conditions in order to cancel the boundary term
in (\ref{eq:10}). However, as we will discuss in the next section, in
the numerical calculations the domain $\Omega$ is always bounded and
hence the boundary conditions (\ref{eq:42}) and (\ref{eq:11}) will be
used.

The procedure to compute the value of $\mf_{min}$ will be the
following. We begin with some arbitrary initial data $(\sigma,
\omega)$ at $t=0$ that satisfies the boundary condition \eqref{eq:27}
for some fixed configuration of $i_k$ and $J_k$.  Then we solve
numerically the flow equations (\ref{eq:haf1})--(\ref{eq:haf2}). The
mass $\mf(t)$ will decrease with time, and it will reach the minimum
value as $t\to \infty$. This minimum will be of course independent of
the initial data. That is, we expect the following behavior of the
solution of the flow equations
\begin{equation}
  \label{eq:13}
  \lim_{t\to \infty}\sigma(t)= \sigma_{min}, \quad\lim_{t\to
    \infty}\omega(t)= \omega_{min},  
\end{equation}
and
\begin{equation}
  \label{eq:14}
  \lim_{t\to \infty} \mf(t)=\mf_{min}.
\end{equation}
Note that (\ref{eq:43}) implies
\begin{equation}
  \label{eq:15}
  \mf(t) \geq \mf_{min} \quad \forall t.
\end{equation}
The monotonicity property (\ref{eq:43}), together with the upper
bound (\ref{eq:15}), make the flow equations ideally suited for a
numerical study of the inequality (\ref{eq:inm}).

Equations (\ref{eq:el1})--(\ref{eq:el2}) are essentially harmonic maps
with singular boundary conditions.  The first existence result for
harmonic maps (in compact manifolds without boundaries) used a heat
flow \cite{eells64}. In that reference the behavior (\ref{eq:13}) was
in fact proved.  There exists also extensions of this result to
include regular boundary conditions \cite{Hamilton75}. These works
were the motivation for the flow equations
(\ref{eq:haf1})--(\ref{eq:haf2}).  We emphasize however, that the
existence results presented in \cite{Dain06c} \cite{Chrusciel:2007ak}
for equations (\ref{eq:el1})--(\ref{eq:haf2}) with the singular
boundary conditions (\ref{eq:27}) (which are based on
\cite{Weinstein96c}) do not use a heat flow, they use a direct
variational method. The numerical calculations presented in this
article confirm (\ref{eq:13}) and hence suggest that a similar
existence result can be proved using the present heat flow.

There is some freedom to construct a heat flow out of equations
(\ref{eq:el1})--(\ref{eq:el2}) in such a way that $\mf$ is monotonic
under the evolution.  Namely we can multiply by arbitrary positive
functions the left hand side of (\ref{eq:haf1})--(\ref{eq:haf2}) and
we still have that $\dot \mf$ is negative. The choice made in
(\ref{eq:haf1})--(\ref{eq:haf2}) appears to be the simplest one because the
principal part of the equations are given by heat equations. In
effect, we can write equations (\ref{eq:haf1})--(\ref{eq:haf2}) as
follows
\begin{align}
  \label{eq:haf1b}
\dot \x  &=  \Delta \x  +\frac{e^{-2\x}|\partial \Y |^2}{\rho ^4},\\
\label{eq:haf2b}
\dot \Y  & = \Delta \Y -4 \frac{\partial_i \omega \partial^i
  \rho}{\rho}-2 \partial_i \omega \partial^i\sigma . 
\end{align}
We also note that in equation (\ref{eq:haf1}) we can apply the maximum
principle for parabolic equations (see, for example, \cite{Evans98}) to
conclude that $\x$ will be positive for all $t$ if the initial data
and boundary conditions are positive.

In this article, the flow (\ref{eq:haf1})--(\ref{eq:haf2b}) is used as
an auxiliary method for computing a solution of the Einstein
stationary equations.  It is however interesting to point out the
relation of this flow with Einstein evolution equations.  As we
mention above, in axially symmetry Einstein equations reduce, in an
appropriate gauge, to a system of wave equations for
$(\sigma,\omega)$. More precisely, these equations have the structure
of ``waves maps'' (see, for example, \cite{tao06} for the definition
of wave maps). The initial conditions for these equations are
essentially the value of $(\sigma,\omega)$ and the value of the time
derivative $(\sigma',\omega')$ on a Cauchy surface. For a typical
collapse initial data, the system will radiate gravitational waves and
reach a final stationary black hole of mass $m_0$. The initial energy
of the system is given by the total mass $m$ and it is conserved along
the evolution (see \cite{Dain:2008xr} for a discussion on this
issue). The total angular momentum $J$ is also conserved along the
evolution.  We always have $m_0\leq m$. These data can be also evolved
using the heat flow. In this case the data are only the value of
$(\sigma,\omega)$ at some time. The total energy of the system if
given by $\mf$, and we have seen that $m\geq \mf$ with equality for
momentary stationary data.  The system will dissipate energy and reach
a final stationary regime with final energy $\mf_{min}$. We have that
$\mf_{min}\leq \mf$.  For the two cases, the system will reach a
solution of the Einstein stationary equations at late time. These
solutions are different, and there is a priori no obvious relation
between them. In particular, there is no obvious relation between
$m_0$ and $\mf_{min}$.

The analogy presented above corresponds essentially to the relation
between wave maps, heat flows and harmonic maps which represents a
geometric generalization of the relation between wave equation, heat
equation and Laplace equation.  For the case without symmetries it is
not possible to reduce Einstein equation to a wave map but the analogy
can still be made if we use the Ricci flow instead of the heat flow.
Note however, that in our case the parabolic equations, although
non-linear, are much simpler than the Ricci flow equations. For a
further discussion about this analogy see \cite{tao06}.


The flow equations will provide a numeric solution $(\sigma,
\omega)$ of equations (\ref{eq:el1})--(\ref{eq:el2}).  As we will
see below, the functions $(\sigma,\omega)$ determine the complete
metric of an stationary axially symmetric spacetime. However,   although the
solution $(\sigma,\omega)$ is always regular outside the ends $i_k$,
it turns out that the other components of the metric are, in general,
not regular at the axis.  That is, not all solutions $(\sigma,\omega)$
will produce a regular spacetime metric.  In particular, it is
expected that a solution $(\sigma,\omega)$ that correspond to many
black holes (i.e. $N\geq 2$ in our setting) do not lead to a regular
metric. As we mentioned in the introduction, this is a relevant point
in the black hole uniqueness theorem.  This is precisely what we
observe in the numerical computations presented in section
\ref{sec:results}.  We emphasize however that in order to test the inequality
(\ref{eq:14a}) we only need to compute the energy $\mf$ which depends
only on $(\sigma,\omega)$ and not on the other components of the
spacetime metric. In particular, the energy $\mf$ is not affected by
the possible singular behavior at the axis of the other components of
the metric.

To reconstruct the spacetime metric from $(\sigma,\omega)$ we follow
\cite{Weinstein90}.  Assume that $(\sigma,\omega)$ are
solutions of equations (\ref{eq:el1})--(\ref{eq:el2}). Then, we can
define, up to constants, the following functions $\Omega$ and
$\gamma$ by
\begin{align}
  \label{eq:16x}
  \gamma_{,\rho} & = \frac{1}{4} \rho \X^{-2}\left(
    \X_{,\rho}^2 -\X_{,z}^2+ \Y_{,\rho}^2 -\Y_{,z}^2  \right),\\
  \gamma_{,z} & = \frac{1}{2} \rho \X^{-2}\left( \X_{,\rho} \X_{,z}+
    \Y_{,\rho}\Y_{,z} \right),
\end{align}
and
\begin{equation}
  \label{eq:6x}
 \Omega_{,z} =\rho \frac{\Y_{,\rho}}{\eta^2},\quad
 \Omega_{,\rho} =-\rho \frac{\Y_{,z}}{\eta^2}.
 \end{equation}
The spacetime metric, in coordinates $(t,\rho,z, \phi)$, is given  by 
\begin{equation}
  \label{eq:24}
  g = -Vdt^2 +2W dt d\phi + \X d\phi^2 + e^{2u}(d\rho^2+dz^2 ),
\end{equation}
where $\eta$ is given in terms of $\sigma$ by \eqref{eq:5} and the
functions $V$, $W$ and $u$ are defined by
\begin{equation}
  \label{eq:17}
  W=\X\Omega, \quad V=X^{-1}(\rho^2-W^2),\quad
  e^{2u}=\frac{e^{2\gamma}}{\X}. 
\end{equation}  
All functions depend only on
$(\rho,z)$. The two  Killing vectors of the metric  are given 
\begin{equation}
\xi^\mu=\left(\frac{\partial}{\partial t} \right)^\mu, \quad
\eta^\mu=\left(\frac{\partial}{\partial \phi} \right)^\mu,
\end{equation}
and we have
\begin{equation}
  \label{eq:53}
V= - \xi^\mu \xi^\nu g_{\mu\nu} , \quad \X = \eta^\mu\eta^\nu
g_{\mu\nu} , \quad
W=\eta^\nu\xi^\mu g_{\nu\mu},   
\end{equation}
where $\mu,\nu$ are spacetime indexes.  We also have that $\Y$ is the
twist potential of $\eta^\mu$ (see \cite{Weinstein90}).

The metric \eqref{eq:24} will be regular at the axis if the 
 following condition is satisfied
\begin{equation}
  \label{eq:18}
\lim_{\rho_0 \to 0^+} \frac{\sqrt{\X}}{\int_0^{\rho_0}e^u d\rho}=1. 
\end{equation}
For arbitrary solutions $(\sigma, \omega)$ this condition will not be
satisfied and hence the metric will not define a regular solutions of
Einstein equations.  The singularities at the axis of these kind of
metrics are interpreted as the forces needed to balance the
gravitational attraction and keep the bodies in equilibrium (see
\cite{Weinstein90} for details).

\begin{figure}
\begin{center}
\includegraphics[width=4cm]{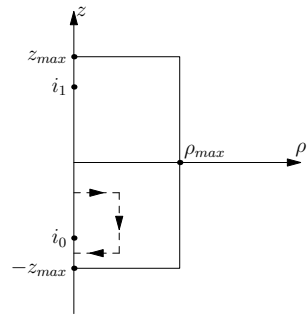}
\caption{The bounded domain for the numerical calculation for two
  black holes located at $i_0$ and $i_1$. The dashed line indicates a
  typical path for the integration of the function $q$.\label{fig:2}}
\end{center}
\end{figure}

The regularity condition \eqref{eq:18} can be conveniently written in term
of a  function $q$ defined by\footnote{In references
 \cite{Weinstein90} \cite{Li92}  a rescaling of this  function is denoted by
 $\beta$, we have the relation $\beta=2q$}
\begin{equation}
  \label{eq:41}
  q=u-\frac{\x}{2}.
\end{equation}
This function satisfies the following equations
\begin{align}
  \label{eq:beta1}
  q_{,\rho} & = \frac{\rho}{4} \left
    ( \x_{,\rho}^2 -\x_{,z}^2\right) +\frac{\rho}{4\X^2}
   \left(\Y_{,\rho}^2- \Y_{,z}^2\right) \\
  q_{,z} & =\frac{1}{2} \rho \left( \x_{,\rho}\x_{,z} +
\X^{-2} \Y_{,\rho}\Y_{,z}  \right). \label{eq:beta2}
\end{align}
Condition \eqref{eq:18} implies that 
\begin{equation}
  \label{eq:47}
  q|_{\rho=0}=0. 
\end{equation}
If the regularity condition fails, we can calculate the value of $q$
at each component of the axis
\begin{equation}
  \label{eq:31}
q_k=q|_{I_k}.
\end{equation}
These values are calculated integrating the gradients
\eqref{eq:beta1}--\eqref{eq:beta2} with an appropriate path, see
figure \ref{fig:2}.  The force between the black holes is given by
\begin{equation}
  \label{eq:30}
F_k=\frac{1}{4}(e^{-q_k} -1).
\end{equation}


Finally, we discuss an important property of the stationary equations
(\ref{eq:el1})--(\ref{eq:el2}), namely their scale invariance (see
\cite{Sudarsky93}).  Let $s>0$ be a real number.  Given functions $\x$
and $\Y$ we define the rescaled functions $\x_s$ and $\Y_s$ by
\begin{equation}
  \label{eq:78}
\x_s(\rho,z)=\x\left(\frac{\rho}{s}, \frac{z}{s}\right), \quad
\Y_s=s^2\Y\left(\frac{\rho}{s}, \frac{z}{s}\right). 
\end{equation}
The functions $(\x_s,\Y_s)$ define solutions of equations
(\ref{eq:el1})--(\ref{eq:el2})  with respect to the rescaled
coordinates $(s\rho, s z)$. Under this scaling, the physical parameters
rescale as
\begin{equation}
  \label{eq:28}
  J \to s^2 J, \quad L_k \to sL_k, 
\end{equation}
and 
\begin{equation}
  \label{eq:82}
  \mf(\x_s,\Y_s)= s\mf(\x,\Y).
\end{equation} 
Note that the quotient $J/L^2$ is scale invariant.  In particular, for
the case of two black holes, with parameters $J_1$, $J_0$ and $L_0$,
the scale invariance of the solution implies that only two parameters
are non-trivial.

\section{The numerical implementation}
\label{sec:Techniques}
We analyze in this section how to solve  numerically the flow
equations \eqref{eq:haf1}--\eqref{eq:haf2}.

\subsection{Equations and boundary conditions}

Although the problem has axial symmetry, equations
\eqref{eq:haf1}--\eqref{eq:haf2} are written in $\Rt$ (the Laplace
operator corresponds to the flat Laplacian in $\Rt$). We can solve
these equations for arbitrary data, with or without axial
symmetry. The minimum will be axially symmetric for any choice of
initial data. This is of course possible because the boundary
conditions (\ref{eq:27}) are axially symmetric. The above
considerations suggest that we can solve numerically the flow
equations in $\Rt$. This approach has the advantage that
no extra boundary conditions on the axis are needed and also that the
equations look more regular in these coordinates.  However, from the
numerical point of view, this method has two major disadvantages. The
first one is that there is a significant loss of resolution because a
3-dimensional grid is used instead of 2-dimensional one.  Second, the
functions are singular at the end points $i_k$ and those points are
inside the domain. That is, there are grid points at both sides of a
singularity and this is very problematic for the finite difference
method.  We found that it is much more convenient to work in a
2-dimensional grid using cylindrical coordinates and  imposing
appropriate boundary conditions on the axis, as we describe in the
following. The only disadvantage of this approach is that we need to
handle terms which are formally singular at the axis. A typical
example is the term $\partial_\rho \sigma /\rho$ which appears in the
cylindrical form of the Laplacian in $\Rt$, namely
\begin{equation}
  \label{eq:8}
  \Delta \sigma= \partial^2_\rho\sigma+ \partial^2_z \sigma+
  \frac{\partial_\rho\sigma}{\rho}. 
\end{equation}
However, following \cite{Garfinkle:2000hd} \cite{Choptuik:2003as},
this kind of terms can be handle numerically in a very satisfactory
manner as we will describe below.

Consider $\Rd$ with coordinates $(\rho,z)$. The domain of interest for
our problem is the half plane $\rho\geq 0$. The axis $\rho=0$ is a
boundary of the domain.   To simplify the notation and the discussion
we will focus on the two black hole problem (i.e. we will have only
two end points $i_0$ and $i_1$ separated by a distance $L$). We
emphasize however that the following discussion trivially extends to
the general case.

In order to handle the singular behavior of the functions at the
points $i_k$ located on the axis, we decompose the solution as
follows.  Let $(\sigma_0, \omega_0)$ be the extreme Kerr solution
  (see  appendix \ref{sec:extr-kerr-solut})  centered at the end $i_0$
with angular momentum $J_0$.  And let $(\sigma_1, \omega_1)$ be the
extreme Kerr solution centered at $i_1$ with angular momentum $J_1$.
Instead of working with $(\sigma,\omega)$, which are singular at
$i_k$, we will work with $(\bar\sigma, \bar\omega)$ defined by
\begin{equation}
  \label{eq:23}
  \x = \x_0 + \x_1+ \bar \x, \quad \Y=\Y_0+\Y_1+\bar \Y.
\end{equation}
The idea is that all the singular behavior of the functions are
contained in $(\sigma_0,\omega_0)$ and $(\sigma_1,\omega_1)$. We
expect the functions $(\bar \sigma, \bar\omega)$ to be regular during
the evolution.

If we insert the ansatz \eqref{eq:23} into the flow equations
\eqref{eq:haf1}--\eqref{eq:haf2} and use the fact that each pair
$(\sigma_0,\omega_0)$ and $(\sigma_1,\omega_1)$ are solutions of the
stationary equations (\ref{eq:el1})--(\ref{eq:el2}), we obtain the
following equations for $(\bar \sigma, \bar\omega)$
\begin{multline}
  \label{eq:haf1d}
\dot{\bar \x}  =  \Delta \bar \x + \frac{e^{-2\x_0}|\partial \Y_0|^2}{\rho ^4}
\left( e^{-2\x_1-2\bar\x} -1 \right)+\\
\frac{e^{-2\x_1}|\partial \Y_1|^2}{\rho ^4} \left( e^{-2\x_0-2\bar\x} -1
\right)+ \frac{e^{-2\x_0-2\x_1-2\bar\x}}{\rho ^4} \left(|\partial\bar
\Y|^2+\right.\\
\left. 2\partial_i  \Y_0 \partial^i \bar\Y + 2\partial_i  \Y_1 \partial^i
\bar\Y +2\partial_i\Y_0 \partial^i \Y_1 \right),
\end{multline}
and
\begin{multline}
\label{eq:haf2d}
\dot{\bar \Y}   =  \Delta \bar \Y  -4 \frac{\partial_i \bar \omega \partial^i
  \rho}{\rho}
-2 \partial_i \bar \omega \partial^i\bar \sigma-
2 \partial_i \bar \omega \partial^i\sigma_0   -2 \partial_i
\bar \omega  \partial^i  \sigma_1-\\
2 \partial_i  \omega_1  \partial^i  
\sigma_0-2 \partial_i\omega_0  \partial^i  \sigma_1-
2 \partial_i  \omega_0  
\partial^i \bar \sigma-2 \partial_i  \omega_1  \partial^i \bar \sigma .
\end{multline}
These are the equations that we actually solve.

Let us analyze the boundary conditions for equations
(\ref{eq:haf1d})--(\ref{eq:haf2d}).  We begin with the axis. The
boundary conditions for the function $\bar \sigma$ at the axis are
given by the regularity conditions. That is, $\bar \sigma$ should be a
regular function in $\Rt$ and hence it should depend smoothly on
$\rho^2$ (see, for example, \cite{Rinne:2005df} \cite{Rinne:2005sk}
for a discussion on regularity conditions at the axis for axially
symmetric problems) and then at the axis it must satisfy
\begin{equation}
  \label{eq:3}
  \partial_\rho \bar \sigma|_{\rho=0}=0 .
\end{equation}
We use equation \eqref{eq:3} as Neumann boundary conditions at the axis. 

For the function $\bar \omega$ the boundary conditions should be such
that they do not change the angular momentum during the
evolution. Since the angular momentum is prescribed by the value of
$\omega$ at the axis, the natural choice is that the angular momentum
is fixed by the values of $\omega_0$ and $\omega_1$ at the axis.
Hence the appropriate boundary condition for $\bar \omega$ at the axis
is the homogeneous Dirichlet one
\begin{equation}
  \label{eq:3b}
   \bar \omega|_{\rho=0}=0.
\end{equation}

If we consider the whole half plane $\rho\geq 0$ as domain, then we
need to prescribe fall off conditions for $\bar\sigma$ and $\bar
\omega$ at infinity compatible with the asymptotic flatness of the
solutions (see \cite{Dain06c}). In particular, the solutions and its
first derivative should go to zero at infinity.

In our case, since the grid is always finite, we need to consider a
bounded domain. The domain will be the rectangle $|z|\leq z_{max}$ and
$0\leq \rho \leq \rho_{max}$, where $z_{max}$ and $\rho_{max}$ are
arbitrary positive constants (see figure \ref{fig:2}).  Let us denote
by $C$ the part of the boundary that does not contain the axis
$\rho=0$.  We need to prescribe boundary conditions on $C$.
These boundary conditions should have two important properties. First,
they should imply that the energy on the domain is monotonic under the
evolution.  Second, in the limit $z_{max}, \rho_{max} \to \infty$ they
should be compatible with asymptotic flatness. That it, in this limit
we want to recover the complete solution on the half plane.

For $\bar \sigma$ we can in principle chose between (\ref{eq:42}) or
(\ref{eq:11}). Note however that an homogeneous Neumann boundary
condition for $\sigma$ translate into an inhomogeneous Neumann
boundary condition for $\bar \sigma$ (since they are related by
equations (\ref{eq:23}) ). Hence, the simpler choice is the 
 homogeneous   Dirichlet condition 
\begin{equation}
  \label{eq:7}
 \bar \sigma|_{C}= 0
\end{equation}
With this choice is also simpler to extend the function to the whole
half plane as we will see below. 

For $\bar \omega$ we can not prescribe Neumann boundary condition on
$C$ since if we do so we can not control the value of $\bar \omega$ at
the points $(0,\pm z_{max})$ where $C$ touch the axis $\rho=0$. In
particular, this will be incompatible with (\ref{eq:3b}). Hence, the only
possibility is to prescribe an homogeneous Dirichlet condition
\begin{equation}
  \label{eq:7b}
 \bar \omega|_{C} =0.
\end{equation}
That is, our set of boundary condition for the numerical evolution is
given by (\ref{eq:3}), (\ref{eq:3b}), (\ref{eq:7}) and (\ref{eq:7b}). 


The variational problem formulated in section
\ref{sec:vari-probl-parab} uses $\Rt$ as domain, which is equivalent,
by the axial symmetry, to the the half plane $\rho \geq 0$.  The fact
that in every numerical computation only a finite domain can be used
will of course introduce an error. In general it is not easy to
measure this error. However, in our case, the variational
characterization of $\mf_{min}$ implies that an upper bound of this
quantity is always obtained even using a finite grid.  This can be
seen as follows. Consider the functions $(\bar \sigma, \bar\omega)$ 
obtained in the numerical evolution of the flow equations
(\ref{eq:haf1d})--(\ref{eq:haf2d}) in the bounded domain
$\Omega$. These functions are, in
principle, only defined in $\Omega$. However, we can extend them to $\Rt$
 imposing that they vanish outside
$\Omega$.  And hence, by (\ref{eq:23}), we get functions $(\sigma,
\omega)$ defined in $\Rt$.  Since $(\bar\sigma,
\bar\omega)$ vanish on $\partial \Omega$ (by the boundary conditions
(\ref{eq:7})--(\ref{eq:7b})) this extension will be continuous but of
course, in general, it will not be differentiable at the boundary
$\partial \Omega$.  However the extension is weakly differentiable. Moreover the
weak derivative is square integrable (see, for example, \cite{Gilbarg}
for the definition of weak derivative and also for the proof of this
fact). Hence, for the extended functions $(\sigma, \omega)$ the
integral \eqref{eq:4} is well defined in $\Rt$ and they satisfy the
boundary conditions (\ref{eq:27}). That is, they
represent admissible test functions for the variational problem. Since
$\mf_{\min}$ is a minimum we have 
\begin{equation}
  \label{eq:1}
  \mf(\sigma, \omega)\geq \mf_{min},
\end{equation}
where we emphasize that in this equation $(\sigma, \omega)$ are the
extended functions. Note that in $\Rt\setminus\Omega$ we have
\begin{equation}
  \label{eq:32}
  \sigma = \sigma_0+ \sigma_1, \quad  \omega=\omega_0+\omega_1, 
\end{equation}
and hence we can decompose the  integral  $\mf(\sigma, \omega)$ as follows
\begin{equation}
  \label{eq:9}
  \mf(\sigma, \omega)= \mf_\Omega(\sigma, \omega)+ \mf_{\Rt\setminus
    \Omega}(\sigma_0 +\sigma_1, \omega_0+\omega_1). 
\end{equation}
The first integral will be the result of the numerical computations
using the heat flow. The second integral depend only on the explicit
functions $(\sigma_0, \sigma_1; \omega_0, \omega_1)$. We computed this 
integral using Maple 9.5.

\subsection{Numerical methods}

We now describe the way we carry out the numerical computations. We use a
finite difference scheme to solve the initial-boundary-value problem (IBVP)
given by the equations (\ref{eq:haf1d})--(\ref{eq:haf2d}) and boundary
conditions (\ref{eq:3})--(\ref{eq:7b}). We also perform numerical integrations
on the computed solutions to evaluate both the mass ${\cal M}_\Omega$ and the
function $q_1$ (see eq. (\ref{eq:31})) used to evaluate the Force
(\ref{eq:30}).

The IBVP is written in cylindrical coordinates $(\rho,z)$ on the domain $0 \le
\rho \le \rho_{max}$ and $-z_{max}\le z\le z_{max}.$ Given two integers
$N_{\rho}$ and $N_z$ we define the step-size in the $\rho$ and $z$ direction
respectively as $h_z=2 z_{max}/N_z$ and $h_\rho=\rho_{max}/N_\rho.$ Our
equations have singular coefficients at the points $i_0$ and $i_1$ on the
$\rho=0$ axis. These point will be placed, in all our runs, at positions $z=
h_z k,$ $k\in{\mathbb Z}.$ The computational grid is defined so that the
gridpoint at the site $(i,j)$ has coordinates
\[\begin{split}
\rho_i &= \Bigl(i - \frac{3}{2}\Bigr) h_{\rho}, \qquad i=0, 1, 2, \dots, N_\rho+2,
N_\rho+3,\\
z_j &= \Bigl(j - \frac{3}{2}\Bigr) h_{z}, \qquad j=0, 1, 2, \dots, N_z + 2,
N_z + 3,
\end{split}
\]
in this way the uniform grid is half a step size displaced with respect to the
coordinate axes and singular points. One can think that the domain is broken
into $N_\rho \times N_z$ cells being each grid point with $2 \le i \le N_\rho +
1$ and $2 \le j \le N_z+1$ placed at the center of a cell. The gridpoints with
$i=0,1,N_\rho+2,N_\rho+3$ and $2\le j\le N_z+1$ are gridpoints at the center of
``ghost cells'' used to impose boundary conditions at the $\rho= \mbox{const.}$
parts of the boundary.  Analogously, the gridpoints with $j=0,1,N_z+2,N_z+3$ and
$2 \le i\le N_\rho+1$ are gridpoints at the center of ``ghost cells'' used to
impose boundary conditions at the $z= \mbox{const.}$ parts of the boundary. The
four gridpoints at each corner of the grid are not used.

\begin{figure}[th]
\psfrag{0}{${}_0$}
\psfrag{1}{${}_1$}
\psfrag{2}{${}_2$}
\psfrag{3}{${}_3$}
\psfrag{4}{${}_4$}
\psfrag{5}{${}_5$}
\psfrag{6}{${}_6$}
\psfrag{7}{${}_7$}
\psfrag{Nr+3}{${}_{N_\rho+3}$}
\psfrag{Nz+3}{${}_{N_z+3}$}
\psfrag{rho}{$\rho$}
\psfrag{z}{$z$}
\psfrag{i}{$i$}
\psfrag{j}{$j$}
\psfrag{to}{$\to$}
\psfrag{up}{$\uparrow$}
\psfrag{dots}{$\dots$}
\includegraphics[width=8cm]{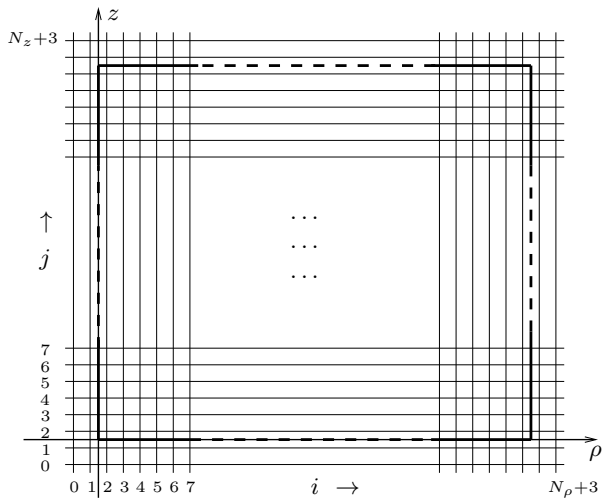}
\caption{Computational grid for the symmetric and antisymmetric cases. The
gridpoints are at the intersection of the thin lines (cells are not shown). 
The rectangle in thick lines is the domain for the IBVP.}
\label{fig:4}
\end{figure}
Many of the problems we actually compute are symmetric or antisymmetric
with respect to the $z=0$ plane. In these cases the symmetry  is used explicitly
to reduce the grid to half-size and so the computer time needed. The grid covers
a domain with $z\ge 0$ (see figure \ref{fig:4}). 

In our numerical scheme, the four partial derivatives $\partial_\rho,$
$\partial_z,$ $\partial^2_\rho$ and $\partial^2_z$ of $\bar \sigma$ and
$\bar\omega$ are approximated by the standard fourth-order accurate symmetric
difference operators \cite{Gustafsson95}
\begin{align}
D &= D_0\Bigl(I - \frac{h}{6} D_+D_-\Bigr), \label{eq:op1}\\
D^2 &= D_+D_-\Bigl(I - \frac{h^2}{12} D_+D_-\Bigr). \label{eq:op2}
\end{align}
Here $D_+$ and $D_-$ denote, as usual, the forward and backward difference
operators, i.e., if $f_i$ is a grid function on a 1-dimensional grid with step
size $h$, we have $D_+ f_i = (f_{i+1} - f_i)/h,$ and $D_- f_i=(f_i -
f_{i-1})/h.$ To be more explicit we show the approximations to the derivatives
with respect to $\rho.$ If $u_{i,j} = u(\rho_i, z_j),$ i.e., $u_{i,j}$ denotes
the grid function associated to the smooth function $u(\rho,z)$, then
\[
\frac{\partial u}{\partial\rho}(\rho_i, z_j) \simeq \frac{\frac{1}{12}
u_{i-2,j} - \frac{2}{3} u_{i-1,j} + \frac{2}{3} u_{i+1,j} - \frac{1}{12}
u_{i+2,j}}{h_\rho},
\]
\begin{eqnarray*}
\lefteqn{\frac{\partial^2 u}{\partial \rho^2}(\rho_i,z_j) \simeq}&& \\
&&\frac{-\frac{1}{12}u_{i-2,j} + \frac{4}{3}u_{i-1,j} -\frac{5}{2}u_{i,j} +
\frac{4}{3}u_{i+1,j} - \frac{1}{12}u_{i+2,j}}{h_\rho^2}.
\end{eqnarray*}

To carry out the time evolution we use the Du Fort-Frankel method. This method
is known to be a good choice for solving parabolic problems because it is
explicit and nevertheless unconditionally stable \cite{Gottlieb:1976} at least
when applied for solving an initial value problem. In the notation of
\cite{Gottlieb:1976} (or \cite[Sect. 7.3]{Gustafsson95}) we set $\gamma=2.$ The
$\gamma$ parameter in this case has to be chosen bigger than $4/3$ for the
method to be stable. The time step can not be chosen big though, and the reason
is twofold. First a big time step gives rise to an increasing
parasitic solution \cite{Gustafsson95} and more important, the boundary
conditions also impose stability restrictions. In the way we treat the boundary
conditions (explained below) we have a scheme that is stable as can be seen
explicitly in our runs, but this scheme is probably not unconditionally stable.
Experimentally we did some runs with a big time step and could see how the
solution diverges in few time steps starting at the boundaries (around the
singular points $i_0$ and $i_1$). In most of our
computations we use $h_\rho = h_z = 10^{-2}$ and a time-step $\delta t =
10^{-4},$ i.e. the square of the space step-size which is the normal ratio in
explicit schemes for parabolic problems.  This time step is however, as the
equations have singular coefficients, around ten times bigger than the time step
we could use with other explicit schemes like 3rd order Runge-Kutta. The Du
Fort-Frankel scheme is only second order accurate but this posses no
inconvenience since we are looking for the stationary solution of the parabolic
problem. In this case, the truncation error due to the time discretization
vanishes when the solution approaches the time independent state.

All the boundary conditions we use are either homogeneous Dirichlet or
homogeneous Neumann boundary conditions. The boundary conditions are imposed to
a grid function via the points at the ghost cells (see for example
\cite{Rinne:thesis}). We show, as example, how this is done for boundary
conditions (\ref{eq:3}) and (\ref{eq:3b}). Given the values of
$\bar\sigma_{i,j}$ and $\bar\omega_{i,j}$ in the interior of the domain, i.e.,
for $2\le i\le N_{\rho}+1$ and $2\le j\le N_z+1$, the values at the ghost cells
with $i=0,1$ are defined as
\[\begin{split}
\bar\sigma_{0,j} &= \bar\sigma_{3,j}, \quad \bar\sigma_{1,j} =
\bar\sigma_{2,j}, \quad \mbox{(Neumann)},\\
\bar\omega_{0,j} &= -\bar\omega_{3,j}, \quad \bar\omega_{1,j} =
-\bar\omega_{2,j} \quad \mbox{(Dirichlet)}.
\end{split}
\]
In this way the boundary conditions (\ref{eq:3}) and (\ref{eq:3b}) are satisfied
exactly to the accuracy order of our computations and the same difference
operators can be used at all gridpoints inside the domain. As we are using the
fourth order accurate operators defined in (\ref{eq:op1}), (\ref{eq:op2}), which
have a span of $\pm 2$ gridpoints, we need two lines of ghost cells outside the
domain for each part of the boundary.

We start the time evolution with with initial data that satisfies the right
boundary conditions.  Now, given the solution at time $t$ satisfying the right
boundary conditions the right hand side of the equations can be computed in the
interior of the domain and the time evolution algorithm computes the values of
$\bar\sigma$ and $\bar\omega,$ in the interior of the domain, at the next time
$t+\delta t.$ Then the solution at this time is extended to the ghost cells so
that it obeys the right boundary conditions and the process is iterated.

Different criteria can be used to stop the time evolution when one is looking
for the stationary state. As the main quantity we want to compute in each run is
the mass ${\cal M}_\Omega$ of the final stationary solution, we stop the run
when the derivative of ${\cal M}_\Omega$ with respect to time becomes, in
absolute value, smaller than a given small value.

To compute the mass ${\cal M}_\Omega$ and the value of $q_1$ we need to
approximate two-dimensional and one-dimensional integrals. As the gridpoints are
placed at the center of cells that cover the domain of our IBVP, the simplest
appropriate rule to approximate these integrals is the midpoint rule.  The
integrand in (\ref{eq:4}), when written in cylindrical coordinates, have singular
points at $i_0$ and $i_1.$ However the midpoint rule provides good enough
results. For example, as in our runs we used vanishing initial data, i.e.,
$\bar\sigma(t=0) = \bar\omega(t=0)= 0,$ the integral in (\ref{eq:4}) becomes an
integral of known, given functions. Thus, we could compare the value obtained
with our code to the value obtained with a very precise integration
rule--implemented in Maple; in the worst case the relative difference between
these values was smaller than $10^{-3}.$

\subsection{Runs and tests}

All the solutions of the IBVP we computed can be divided into three groups. The
first group consists of symmetric configurations in which $J_0=-1.0$ is placed at
$z=-L/2$ and $J_1 = J_0$ placed at $z=L/2.$ Within this group we carried out
runs for different values of $L,$ and for different domain sizes. It is clear 
in this case the solutions $\sigma$ and $\omega$ of
(13)--(14) are, respectively, symmetric and antisymmetric as functions of $z.$
Moreover $\bar\sigma$ and $\bar\omega$ satisfy the same symmetry
during the whole time evolution of our IBVP---even on a finite domain---provided
the domain itself and the initial data are symmetric. The obvious initial data
satisfying all boundary conditions and symmetry is $\bar\sigma(t=0) =
\bar\omega(t=0) = 0;$ this is what we used in all our runs.

The second group of solutions we computed correspond to antisymmetric
configurations in which we placed $J_0=-1.0$ at $z=-L/2$ and $J_1=-J_0$ at
$z=L/2.$ We carried out runs for different values of $L.$
The solutions in this group also have a clear symmetry. In
this case both $\bar\sigma$ and $\bar\omega$ are symmetric as functions of $z.$

The third group of solutions we computed correspond to asymmetric
configurations in which we placed $J_0=1.0$ at $z=-1/2$ and $J_1 \neq J_0$ at
$z=1/2.$ Within this group we carried out runs for various values of $J_1.$

When computing solutions in the symmetric or antisymmetric configurations we
need to compute the solution in half the domain only, $z\in[0,z_{max}]$ and
$\rho\in[0,\rho_{max}].$ $z=0$ becomes a boundary and all we need is to use
extra boundary conditions at $z=0$ that obey the symmetry of the solutions.
This boundary conditions are homogeneous Neumann for $\bar\sigma$ and
homogeneous Dirichlet for $\bar\omega$ in the symmetric case, and homogeneous
Neumann for both functions in the antisymmetric case. By using the symmetry of
the solution we reduce to one half the computer time needed.

A main issue, from the point of view of the numerical
calculations, is to determine the size of the domain where to compute ${\cal
M}_\Omega.$ At the same time we need to estimate error we commit in the determination 
of ${\cal M}.$ We attack these questions mainly by studying the symmetric case.
 
\begin{figure}[tp]
{\scriptsize
\psfrag{L01}{$0.1$}
\psfrag{L1}{$1$}
\psfrag{L2}{$2$}
\psfrag{L1}{$1$}
\psfrag{L2}{$2$}
\psfrag{L3}{$3$}
\psfrag{L4}{$4$}
\psfrag{L5}{$5$}
\psfrag{L6}{$6$}
\psfrag{L7}{$7$}
\psfrag{L8}{$8$}
\psfrag{MOmega}{${\cal M}_\Omega$}
\psfrag{t}{$t$}
\includegraphics[width=8cm]{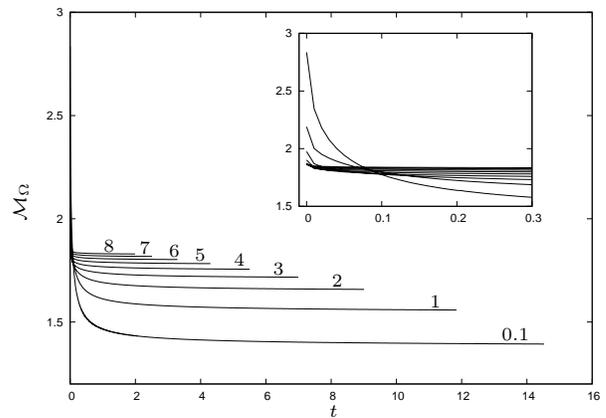}}
\caption{Time evolution of ${\cal M}_\Omega$ for different values of $L$ (number
close to each curve). In this plot, all the runs were stopped when $|d{\cal
M}_\Omega/dt|<5.0\times10^{-4}$. The detail shows the evolution for a short while
$t\in[0,0.3].$}\label{fig:evolution}
\end{figure}

The time evolution of ${\cal M}_ \Omega$, for different values of $L$, can be
seen in figure \ref{fig:evolution}. The initial data in all the runs was set to
zero. The smaller the value of $L$ is the bigger the initial ${\cal M}_\Omega$ is,
and also the stronger the equations dissipate so that the code runs for a
longer time and the final ``stationary'' ${\cal M}_\Omega$ turns out to be
smaller.

With the purpose of evaluating the precision of the values of mass obtained and
of determining a convenient domain size to carry out our computations we performed
runs with the same physical parameters but on different domains (and
corresponding grids). The results are
shown in table \ref{table:domains} for the two smallest values of $L$.
\begin{table*}[htp]
\begin{tabular}{|c|c|c|c|c|c|c|}
\hline
Domain $(\rho,z)$ & ${\cal M}_{\Omega 0}$ & ${\cal M}^M_{\Omega 0}$ & Rel. error & ${\cal
M}_\Omega$ & ${\cal M}_0$ & ${\cal M}$ \\
\hline
$[0,10]\times[-5,5]$   & 2.650041128 & 2.651146040 & $-4.17\times 10^{-4}$ &
1.220646770 & 2.898066024 & 1.467566754 \\
$[0,20]\times[-10,10]$ & 2.773619709 & 2.774724666 & $-3.98\times 10^{-4}$ &
1.332235504 & 2.898066024 & 1.455576862 \\
$[0,40]\times[-20,20]$ & 2.835391328 & 2.836496292 & $-3.90\times 10^{-4}$ &
1.393365251 & 2.898066024 & 1.454934983 \\
$[0,80]\times[-40,40]$ & 2.866221110 & 2.867326074 & $-3.85\times 10^{-4}$ &
1.424331374 & 2.898066024 & 1.455071324 \\
\hline                                                
$[0,10]\times[-5,5]$   & 2.002782916 & 2.003994940 & $-6.05\times 10^{-4}$ &
1.381272815 & 2.251736983 & 1.629014858 \\
$[0,20]\times[-10,10]$ & 2.127077532 & 2.128289601 & $-5.70\times 10^{-4}$ &
1.496600061 & 2.251736983 & 1.620047443 \\
$[0,40]\times[-20,20]$ & 2.188941879 & 2.190153954 & $-5.53\times 10^{-4}$ &
1.558251402 & 2.251736983 & 1.619834431 \\
$[0,80]\times[-40,40]$ & 2.219783296 & 2.220995372 & $-5.46\times 10^{-4}$ &
1.589210683 & 2.251736983 & 1.619952294 \\
\hline
\end{tabular}
\caption{Several runs with the symmetric configuration and the same physical
parameters, $J_0 = J_1 = 1.0,$ but on different domains.
In all cases $h_\rho = h_z = 10^{-2},$ $\delta t = 10^{-4}$ and the
initial data was set to zero. The upper half of the table corresponds to $L=0.1$
and the lower part to $L=1.0$}\label{table:domains}
\end{table*}
In table \ref{table:domains} ``${\cal M}_{\Omega 0}$'' is the value of ${\cal
  M}_\Omega(t=0)$ for vanishing initial data ($\bar\sigma(t=0) =
\bar\omega(t=0)=0$) as computed by our program; ``${\cal M}^M_{\Omega 0}$'' is
the same quantity as computed by an integration routine of Maple 9.5.  ``${\cal
  M}_\Omega$'' is the value computed by our program when the solution is close
enough to the stationary state ($|d{\cal M}_\Omega/dt|<5.0\times 10^{-4}$ for
these runs). ``${\cal M}_0$'' is the value of the total initial energy,
computed with Maple 9.5, on a huge domain
$(\rho,z)\in[0,40000]\times[-20000,20000].$ Finally ``${\cal M}$'' is given by
${\cal M}= {\cal M}_\Omega + ({\cal M}_0 - {\cal M}^M_{\Omega 0}).$ On the one
hand we have the error introduced by the integration routine.  Comparing the
second and third columns of the table we see that our integration routine can
guarantee three correct figures (two after the decimal point) at initial
time. We assume this also holds at final time. On the other hand there is the
error introduced by the compactness of the computational domain. Each domain
used quadruples the previous domain in size. The values of ${\cal M}$ obtained
for the three largest domains are coincident when we round the figures to four
digits. Based on this facts we are confident enough as to choose the domain
$(\rho,z)\in[0,40]\times[-20,20],$ in which our code runs fast enough, for all
our computations. Hence, we accept as correct the computed values of ${\cal M}$
rounded to three digits. The same domain was used to perform the runs in the
antisymmetric and asymmetric cases.

\section{Results}\label{sec:results}

In this section we present the results of the numerical
simulations. As we pointed out above, we will concentrate on the two
black holes case with individual momentum $J_0$,  $J_1$ and separated
by a distance $L$.  

 \subsection{Expected behavior}
\label{sec:expected-behavior}
Let us first discuss, in an heuristic way, the expected behavior of
the total mass $\mf_{min}$ of the stationary solution corresponding to
this configuration in some asymptotic limits.

Consider the far limit $L\to \infty$. In this limit, we expect the
interaction between the black holes to be small. If we make an
expansion in powers of $L^{-1}$ of the total mass $\mf_{min}$ the
first non-trivial terms should correspond to the sum of the individual
masses. The second term should correspond to the Newtonian
gravitational interaction energy between the black holes. And finally,
the third term should be given by the interaction between the angular
momentum of the black holes. This term is called spin-spin
interaction in the literature (see \cite{Wald72} \cite{Dain02e} for a
detailed discussion of this issue). That is, we expect the following
expansion
\begin{equation}
  \label{eq:46}
  \mf_{min}\approx m_0+m_1- \frac{m_0 m_1}{L}+ \frac{2J_0 J_1}{L^3}+O(L^{-4}),
\end{equation}
where $m_0=\sqrt{|J_0|}$ and $m_1=\sqrt{|J_1|}$. This kind of
expansion is valid without the assumption of axial symmetry, in fact
the formula (\ref{eq:46}) arises as particular case of the general
expansion presented in \cite{Wald72} \cite{Dain02e}. 
Also, for the solution $(\sigma_{min}, \omega_{min})$ we expect the
following behavior in the limit $L \to \infty$
\begin{equation}
  \label{eq:20}
\sigma_{min}\approx \sigma_0+\sigma_1,\quad  \omega_{min}
\approx\omega_0+\omega_1.   
\end{equation}
The Newtonian interaction is of course always negative. However the sign of the
spin-spin interaction depends on the individual signs of the angular momentum
$J_0$ and $J_1$. For the aligned case (i.e. when $J_0$ and $J_1$ has the same
sign) it is positive and hence has opposite sign as the Newtonian
interaction. This is the most interesting case regarding the inequality
(\ref{eq:14a}) since it is expected to be the most favorable situation to find
a counter example to conjecture \ref{c:1}. This can be seen as follows. In a
configuration with aligned angular momentum the total amount of angular
momentum $|J|$ (recall that $J=J_0+J_1$) is always greater than in the anti
aligned case. On the other hand the first terms in the expansion (\ref{eq:46})
are the same in both cases. That is, up to order $L^{-2}$ we have the same mass
in both configurations but the aligned one has greater total angular
momentum. Also, from the point of view of the black hole equilibrium problem
the aligned configuration is the most interesting one since in this case it is
in principle conceivable that the spin-spin force balance the Newtonian
gravitational attraction to make the equilibrium possible at some particular
separation distance $L$. On the other hand, for the anti aligned case it has
been proved that the equilibrium is not possible \cite{Li91b}.

Let us discuss now the limit $L\to 0$. In this limit we have only one
asymptotic end and hence we expect that the solution approach to a
single extreme Kerr solution with angular momentum $J$. Let us denote
this solution by $(\sigma_{01}, \omega_{01})$. This behavior can be
justified as follows.  Consider the behavior of the individual
extreme Kerr solutions $(\sigma_0, \omega_0)$ and
$(\sigma_1,\omega_1)$. The sum $(\sigma_0+ \sigma_1 ,
\omega_0+\omega_1)$ it is of course not a solution of the stationary
equations (\ref{eq:el1})--(\ref{eq:el2}) even in this limit since the
equations are not linear. However, the extreme Kerr solutions have a
`linear piece' namely the functions $\hat \omega_0$ and $\hat
\omega_1$ which fix the angular momentum of the solution (see
equation (\ref{eq:57c})). In the limit $L\to 0$ this sum corresponds
to $\hat \omega_{01}$. Since this is the part of $\omega$ that fixes
the angular momentum and the solution is unique for fixed angular
momentum, the flow equation should produce functions $(\bar\sigma,
\bar\omega)$ such that the final $(\sigma,\omega)$ approach to the
Kerr extreme solution with angular momentum $J$. That is, in the limit
$L\to 0$ we expect the following behavior
\begin{equation}
  \label{eq:6}
   \mf\approx \sqrt{|J|}, 
\end{equation}
and
\begin{equation}
  \label{eq:21}
\sigma_{min}\approx \sigma_{01},\quad  \omega_{min} \approx\omega_{01}.  
\end{equation}
In this limit the function $\bar\sigma$ is singular. This can be seen
as follows.  In the limit $L\to 0$, by equation (\ref{eq:25}), the
singular behavior of the sum $\sigma_0+\sigma_1$ is given by
\begin{equation}
  \label{eq:26}
  \sigma_0+\sigma_1= -4\log r +O(1).
\end{equation}
On the other hand, by the same equation (\ref{eq:25}), the behavior
of the extreme Kerr solution $\sigma_{01}$ is given by
\begin{equation}
  \label{eq:29}
  \sigma_{01}=-2\log r +O(1).
\end{equation}
Then, in order to reach (\ref{eq:29}), the function $\bar \sigma$
generated by the flow should be singular in the limit $L\to 0$. This
is precisely what we observe in our computation as we will see.

Finally, let us discuss the general shape of the curve
$\mf_{min}(L)$. This curve should not have minimum or maximum, since,
roughly speaking, a minimum or maximum will signal an equilibrium
point. Using the asymptotic limits (\ref{eq:46}) and (\ref{eq:6}), we
conclude that the curve $\mf_{min}(L)$ should lie between the lines
$\sqrt{|J|}$ and $\sqrt{|J_0|}+\sqrt{|J_1|}$ and it should be
  monotonically increasing with $L$, that is
  \begin{equation}
    \label{eq:33}
    \frac{\partial\mf_{min}(L) }{\partial L} >0.
  \end{equation}

\subsection{Results of computations}


Let us consider first the symmetric configuration, that is, two black holes with
the same angular momentum $J_0=J_1=J$ separated by a distance $L$. As we have
discussed in section \ref{sec:vari-probl-parab}, by the scale invariance of the
equations, we can normalize to $J=1$ without loss of generality. 

\begin{figure}[bhp]
{\scriptsize
\psfrag{z}{$z$}
\psfrag{rho}{$\rho$}
\psfrag{bs}{$\!\!\bar\sigma$}
\includegraphics[width=8cm]{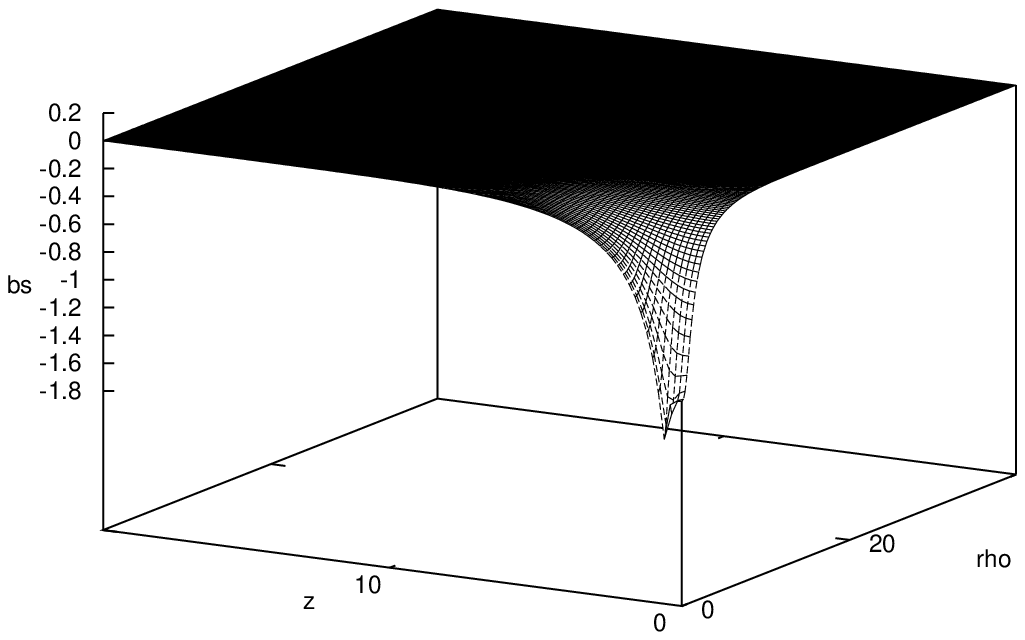}
\includegraphics[width=8cm]{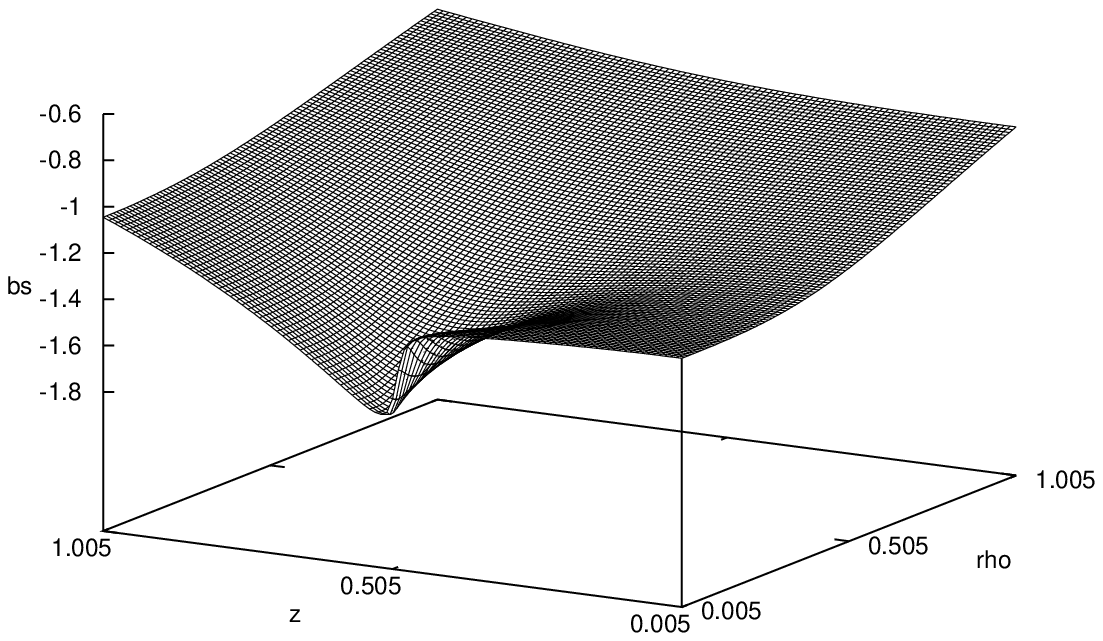}
}
 \caption{Plots of $\bar\sigma$ in the semi-domain $z\in[0,20],~\rho\in[0,40]$
 (above) and detail of the same graph in a small square region $z\in[0.005,
 1.005]$ and $\rho\in[0.005, 1.005]$ to show the behavior of the solution close
 to the singular point $i_1$ (located at $\rho=0$ and $z=0.05$).
 }\label{fig:sigmabar}
 \end{figure}

\begin{figure}[bhp]
 {\scriptsize
 \psfrag{z}{$z$}
 \psfrag{rho}{$\rho$}
 \psfrag{bo}{$\!\!\bar\omega$}
 \includegraphics[width=8cm]{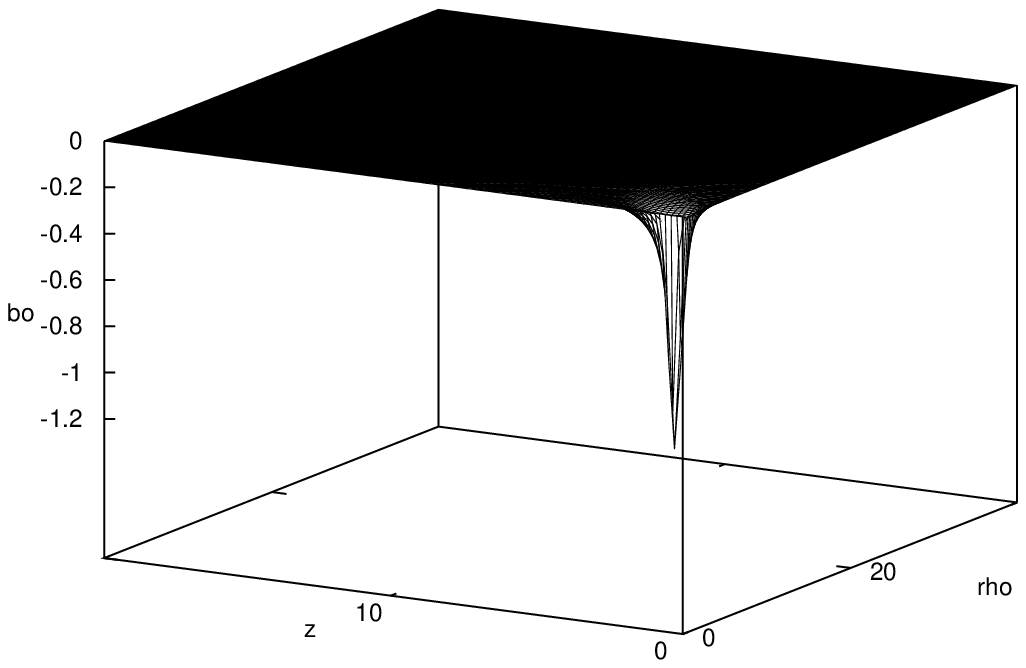}
 \includegraphics[width=8cm]{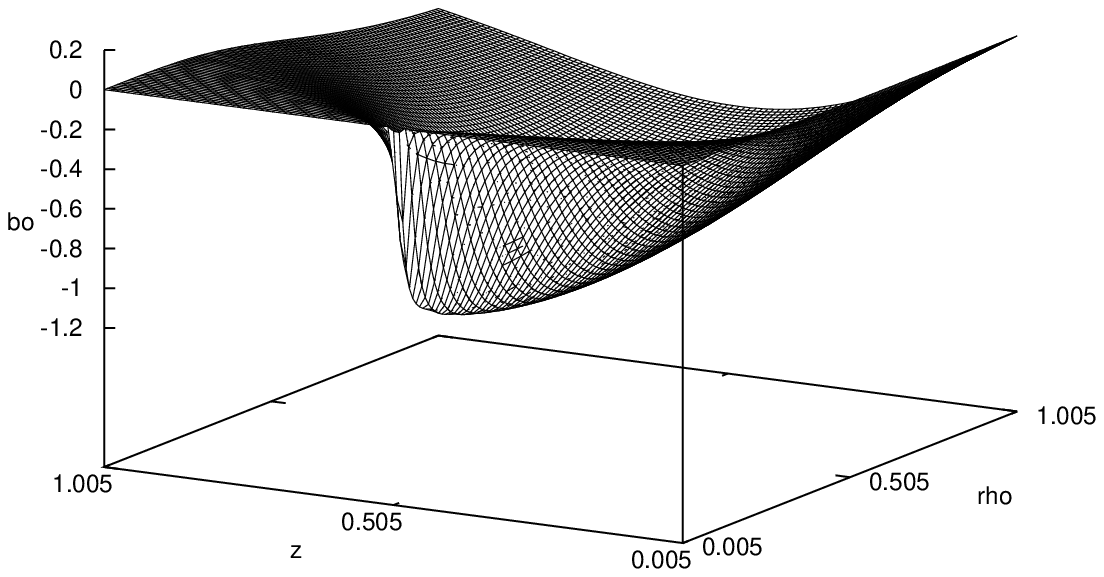}
 }
 \caption{Plots of $\bar\omega$ in the semi-domain $z\in[0,20],~\rho\in[0,40]$
 (above) and detail of the same graph in a small square region $z\in[0.005,
 1.005]$ and $\rho\in[0.005, 1.005]$ to show the behavior of the solution close
 to the singular point $i_1$ (located at $\rho=0$ and
 $z=0.05$).}\label{fig:omegabar}
 \end{figure}

In figures \ref{fig:sigmabar} and \ref{fig:omegabar} we present the plots of
$\bar\sigma(\rho,z)$ and $\bar\omega(\rho,z)$ for the symmetric case with $L=1$
in the semi-domain as a typical plot of the solutions obtained. A detail of
each plot near the ends, where most of the variations of the functions occur,
is also shown.

\begin{figure}[hbp]
 \begin{center}
 {\scriptsize
 \psfrag{sr2}{$\!\!\!\!\sqrt{2}$}
 \psfrag{L}{$L$}
 \includegraphics[width=8cm]{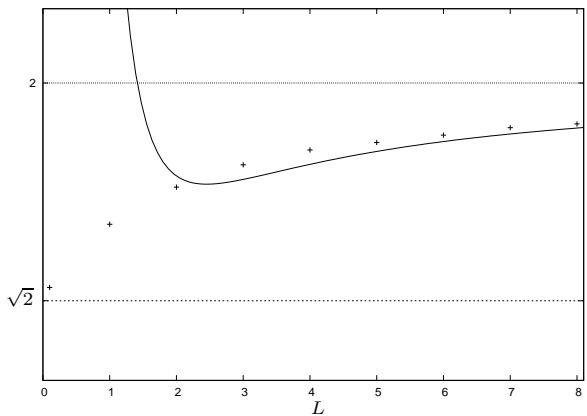}
 }
 \caption{A plot of the values presented in table \ref{table:symmetric}. The
 individual points are the values of ${\cal M}.$ We have
 also plotted the Newtonian interaction plus the spin-spin interaction
 given by equation (\ref{eq:46}). The two horizontal lines located at
 $2$ and $\sqrt{2}$ indicate the sum of the individual masses
 $m_0+m_1=2$ and the total angular momentum $J=\sqrt{2}$
 respectively.}\label{fig:main_plot}
 \end{center}
 \end{figure}

 Our main result is shown in table \ref{table:symmetric} where we present the
 computed values of ${\cal M}$ (rounded to three digits).
 \begin{table}
 \begin{tabular}{|c|c|c|c|}
 \hline
 $L$ & ${\cal M}_\Omega$ initial & \hspace{0.2em}${\cal M}_\Omega$
 final\hspace{0.2em} & \hspace{1.5em}${\cal M}$\hspace{1.5em} \\
 \hline 
 0.1 & 2.84 & 1.39 & 1.45 \\
 1.0 & 2.19 & 1.56 & 1.62 \\
 2.0 & 1.98 & 1.66 & 1.72 \\
 3.0 & 1.90 & 1.72 & 1.78 \\
 4.0 & 1.88 & 1.76 & 1.82 \\
 5.0 & 1.87 & 1.78 & 1.84 \\
 6.0 & 1.87 & 1.80 & 1.86 \\
 7.0 & 1.87 & 1.82 & 1.88 \\
 8.0 & 1.87 & 1.83 & 1.89 \\
 \hline
 \end{tabular}
 \caption{Computed values of ${\cal M}_\Omega$ and final energy ${\cal M}$ for
   different values of $L$ in the symmetric configuration. 
   The individual angular momentum parameters are $J_0=J_1=1$, and hence we have
   $\sqrt{|J|}=\sqrt{2}$.
   The domain used was
   defined by $z_{max}=20$ and $\rho_{max}=40$. The grid used (for the semi-domain)
   is $4000\times2000$ points.}\label{table:symmetric}
 \end{table}
 Figure \ref{fig:main_plot} shows the plot ${\cal M}$ as function of $L.$
 Clearly all values of ${\cal M}$ are higher than $\sqrt{2}$ and conjecture
 \ref{c:1} is satisfied. Direct observation of figure
 \ref{fig:main_plot} shows that ${\cal M}$ is a monotonic function of $L$ and, at
 least graphically, seems to obey that ${\cal M}\to \sqrt{2}$ when $L\to 0.$
 Moreover, although the values of $L$ for which we could compute the solution are not
 big, the plot also shows that the limit ${\cal M}\to 2$ when $L\to \infty$ is
 plausible.

 For every value of $L$ in table \ref{table:symmetric} we also computed the force
 between the black-holes given by equation (\ref{eq:30}) and the value of the
 derivative $d{\cal M}/dL.$ To compute the force we evaluated the corresponding
 value of $q_1$ by following ten different trajectories surrounding the end
 $i_1.$ We found that the values of $q_1,$ for the different trajectories, are
 not the same as expected. The reason is that $q_1$ is a much more sensitive
 measurement of ``stationarity'' of the solution than ${\cal M}$ is and so we
 would need a much longer evolution to have a good value of $q_1$ (see
 explanation below). All the values of $q_1$ have the right sign though, and the
 force between the black holes is always attractive. Thus, we present in table
 \ref{table:force} coarse values of the force. It is important to stress, though,
 that the sign is correct and the value of the force is decreasing with $L$ in
 coincidence with the values of the derivative of ${\cal M}.$ Shown values of
 $d{\cal M}/dL$ were obtained by computing values of ${\cal M}$ at two extra
 nearby values of $L+\delta L$ and approximating the derivative by a symmetric
 finite difference operator.

 Naively, we would expect that the force $F$ is equal to the derivative $d{\cal
   M}/dL$. From our data, we observe that this equality appears only to be true
 in the limit $L\to \infty$. However, since our values for $F$ are coarse, this
 difference could in principle be a consequence of numerical errors.

 \begin{table}
 \begin{tabular}{|c|c|c|c|}
 \hline
 $L$ & \hspace{1.5em}$q_1$\hspace{1.5em} & \hspace{1.5em}$F_1$\hspace{1.5em} &
 \hspace{0.2em}$d{\cal M}/dL$\hspace{0.2em} \\
 \hline
 0.1 & -1.00  & 0.430 & 0.247 \\
 1.0 & -0.53  & 0.174 & 0.133 \\
 2.0 & -0.30  & 0.088 & 0.074 \\
 3.0 & -0.20  & 0.054 & 0.047 \\ 
 4.0 & -0.14  & 0.038 & 0.032 \\ 
 5.0 & -0.11  & 0.028 & 0.023 \\ 
 6.0 & -0.082 & 0.021 & 0.017 \\ 
 7.0 & -0.067 & 0.017 & 0.013 \\ 
 8.0 & -0.052 & 0.013 & 0.011 \\
 \hline
 \end{tabular}
 \caption{Values of $q_1,$ the attractive force between the black holes and
 derivative of ${\cal M}$ w.r.t. $L.$}\label{table:force}
 \end{table}
 Using a small domain, so that our code runs fast, we performed two runs with the
 same physical parameters ($L=1$) but stopping the time evolution with two
 different criteria. The short run stopped when $d{\cal M}/dt <5.0\times 10^{-4}$
 as most of our runs. The long run stopped when the both the absolute values of
 the time derivatives of $\bar \sigma$ and $\bar\omega,$ at every site, was
 smaller than $10^{-5},$ i.e., this last criterion sensed stationarity pointwise.
 When the long run stopped, the value of $d{\cal M}/dt$ was around $10^{-10}$ but
 the value of ${\cal M}$ itself was coincident up to four digits to that of the
 short run. The values of $q_1$ computed at final time on ten different
 trajectories around $i_1$ for both runs showed a variation of $1.27\%$ for the
 short run and $0.025\%$ for the long run.

\begin{figure}[th]
 {\scriptsize
 \psfrag{sigmabar}{$\bar\sigma$}
 \psfrag{z}{$z$}
 \includegraphics[width=8cm]{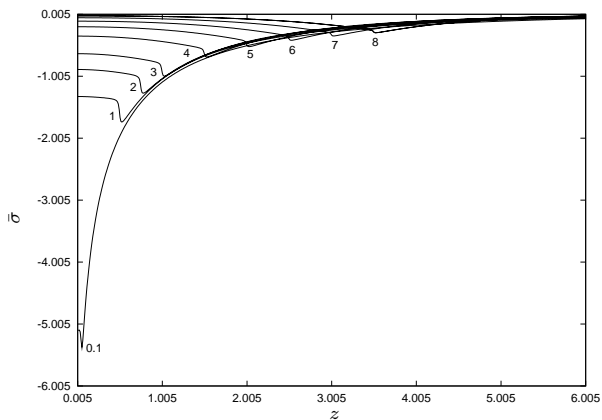}
 }
 \caption{Plots of $\bar\sigma$ as functions of $z$ at $\rho=h_\rho/2$ (the
 closest gridpoints to the symmetry axis) for different values of $L$ (numbers
 close to each curve). Only the $z>0$ part of the plot is
 shown.}\label{fig:picos_sigma_bar}
 \end{figure}

 The function $\bar\sigma$ is bounded on the whole domain. However, one can see
 how the peak values of sigma at the symmetry axis, $\bar\sigma(\rho=0),$ occur
 very close or at the singular points $i_0$ and $i_1.$ In the symmetric case
 one, as we have seen in section \ref{sec:expected-behavior}, we expect that
 the value of $\bar\sigma$ in the axis, in the region between the ends $i_0$
 and $i_1$ diverges as $L\to 0.$ We could observe this behavior in our
 numerical solutions. Figure \ref{fig:picos_sigma_bar} shows the plots of
 $\bar\sigma(\rho=h_\rho/2,z)$ as function of $z.$ The expected divergent
 behavior as $L\to 0$ is clearly seen in the graph.

 We consider now the anti-symmetric case, where $J_0=-J_1$ and separated by a
 distance $L$. Since the total angular momentum in this case is zero, the
 inequality (\ref{eq:inm}) is trivially satisfied. Nevertheless, it is
 important to compute this case for testing purpose.

\begin{figure}[hbp]
 {\scriptsize
 \psfrag{L}{$L$}
 \psfrag{M}{${\cal M}$}
 \includegraphics[width=8cm]{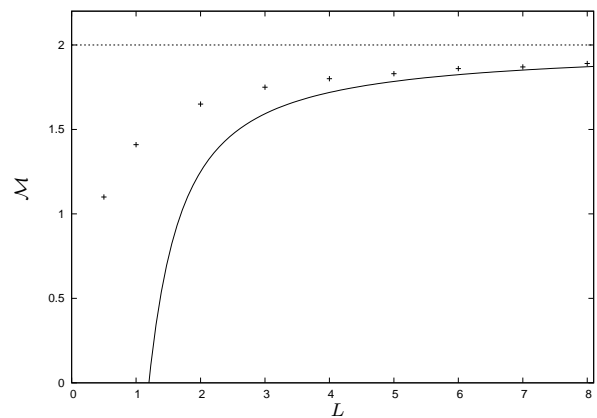}
 }
 \caption{Total mass in the anti-symmetric case as function of
 $L.$ $J_0 = -1$ is located at $z=-L/2$ and $J_1=1$ is located at $z=L/2.$ The
 semi-domain is $(\rho,z) \in [0,40]\times[0,20].$ The continuous line is the
 Newtonian plus spin-spin interaction.}\label{fig:anti_plot}
 \end{figure}

 As in the previous case we fixed the angular momentum and normalize them to
 $J_0=1$ and $J_1=-1$. The results obtained are shown in table
 \ref{table:antisymmetric} and plotted in figure \ref{fig:anti_plot}.
 \begin{table}[htp]
 \begin{tabular}{|c|c|c|c|}
 \hline
 $L$ & ${\cal M}_\Omega$ initial & \hspace{0.2em}${\cal M}_\Omega$
 final\hspace{0.2em} & \hspace{1.5em}${\cal M}$\hspace{1.5em} \\
 \hline 
 0.5 & 2.39 & 1.04 & 1.10 \\
 1.0 & 2.14 & 1.35 & 1.41 \\
 2.0 & 1.95 & 1.59 & 1.65 \\
 3.0 & 1.88 & 1.68 & 1.75 \\
 4.0 & 1.86 & 1.74 & 1.80 \\
 5.0 & 1.86 & 1.77 & 1.83 \\
 6.0 & 1.86 & 1.79 & 1.86 \\
 7.0 & 1.86 & 1.81 & 1.87 \\
 8.0 & 1.86 & 1.82 & 1.89 \\
 \hline
 \end{tabular}
 \caption{Computed values of ${\cal M}_\Omega$ and final energy ${\cal M}$ for
   different values of $L$ in the antisymmetric configuration $J_0=-J_1=1$, where the total
   angular momentum $J$ is zero.  The domain used was
   defined by $z_{max}=20$ and $\rho_{max}=40$. The grid used (for the semi-domain)
   is $4000\times2000$ points.}\label{table:antisymmetric}
 \end{table}

 Finally, we consider the asymmetric case, where $J_0$ and $J_1$ are separated
 by a distance $L=1.$ We perform runs with $J_0=-1$ and varying $J_1 \in [-1.0,
 1.0].$ This case is interesting because in the limit $J_1=0$ we must recover
 the one extreme Kerr black hole solution and hence the equality in
 (\ref{eq:inm}).  Note that this limit (as the limit $L\to 0$ in the symmetric
 case) is a singular limit. In both limits we are exploring the neighborhood of
 the equality in (\ref{eq:inm}) and hence the most favorable cases for a
 possible counter example. 

\begin{figure}[hbp]
 {\scriptsize
 \psfrag{J1}{$J_1$}
 \psfrag{M}{${\cal M}$}
 \includegraphics[width=8cm]{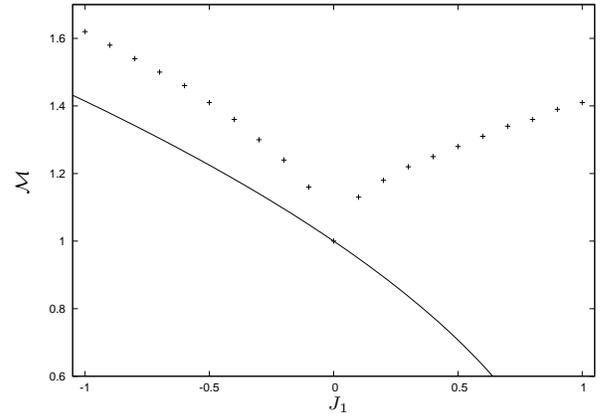}
 }
 \caption{Total mass in the asymmetric case as function of
 $J_1.$ $J_0 = -1$ is located at $z=-1/2$ and $J_1$ is located at $z=1/2.$ The
 semi-domain is $(\rho,z) \in [0,40]\times[0,20].$ The continuous line is the
 lower bound according to the conjecture \ref{c:1}}\label{fig:asym_plot}
 \end{figure}

 The results are shown in table \ref{table:asymmetric} and plotted in figure
 \ref{fig:asym_plot}. We observe that the inequality  (\ref{eq:inm}) is
 satisfied in all cases.  
 \begin{table}
 \begin{tabular}{|c|c|c|c|c|}
 \hline
 $J_1$ & ${\cal M}_\Omega$ initial & \hspace{0.2em}${\cal M}_\Omega$
 final\hspace{0.2em} & \hspace{1.5em}${\cal M}$\hspace{1.5em} &
 $\sqrt{|J|}$  \\
 \hline 
 -1.0 & 2.19 & 1.56 & 1.62 & 1.41  \\
 -0.9 & 2.12 & 1.52 & 1.58 &1.38  \\
 -0.8 & 2.06 & 1.49 & 1.54 & 1.34 \\
 -0.7 & 1.98 & 1.45 & 1.50 &1.30 \\
 -0.6 & 1.91 & 1.41 & 1.46 &1.26 \\
 -0.5 & 1.82 & 1.36 & 1.41 & 1.22 \\
 -0.4 & 1.72 & 1.32 & 1.36 & 1.18 \\
 -0.3 & 1.61 & 1.27 & 1.30 &1.14  \\
 -0.2 & 1.48 & 1.21 & 1.24 & 1.10 \\
 -0.1 & 1.32 & 1.13 & 1.16 & 1.05 \\
  0.0 & 0.98 & 0.98 & 1.00 & 1 \\
  0.1 & 1.30 & 1.10 & 1.13 & 0.95 \\
  0.2 & 1.46 & 1.15 & 1.18 & 0.89 \\
  0.3 & 1.59 & 1.18 & 1.22 & 0.84 \\
  0.4 & 1.69 & 1.21 & 1.25 & 0.77 \\
  0.5 & 1.78 & 1.24 & 1.28 & 0.71 \\
  0.6 & 1.87 & 1.24 & 1.31 &0.63  \\
  0.7 & 1.94 & 1.28 & 1.34 & 0.55 \\
  0.8 & 2.02 & 1.30 & 1.36 & 0.45 \\
  0.9 & 2.08 & 1.33 & 1.39 & 0.32 \\
  1.0 & 2.14 & 1.35 & 1.41 & 0 \\
 \hline
 \end{tabular}
 \caption{Computed values of ${\cal M}_\Omega$ and final energy ${\cal M}$ in the
 asymmetric configuration as function of $J_1$, where $J_0=-1$. The separation
 distance is  $L=1$ and the location of $i_0$  is fixed at $z=-0.5$
 and $i_1$ is fixed at $z=0.5.$ The domain used was defined by $z_{max}=20$
 and $\rho_{max}=40$. The grid used is $4000\times4000$
 points.}\label{table:asymmetric}
 \end{table}

 \section{Conclusion}\label{sec:conclusion}
 The main result of this article is given in tables \ref{table:symmetric} and
 \ref{table:asymmetric}. In all cases, we have verified the inequality
 (\ref{eq:inm}). That is, we have provided strong numerical evidence that this
 inequality is true for two axially symmetric black holes. Moreover, we have
 computed a non zero force in the symmetric case (table \ref{table:force}) and
 hence we have also provided numerical evidences that the equilibrium is not
 possible for two extreme black holes.

 The monotonic dependence of the total energy $\mf$ in terms of the separation
 distance $L$ plotted in figure \ref{fig:main_plot} suggests a possible
 strategy to prove analytically the inequality (\ref{eq:inm}). Namely, to study
 the neighborhood of $L=0$ of the energy. In particular, the first step is to
 prove that $d\mf/dL >0 $ at $L=0$. Since the value of $\mf$ at $L=0$ is known,
 this will prove the inequality near $L=0$.  The second step (probably much
 more difficult) will be to prove that $d\mf/dL >0 $ for any $L$.

 Finally, we have also shown that the heat flow equations
 (\ref{eq:haf1})--(\ref{eq:haf2}) constitute an efficient and simple numerical
 method to construct solutions of the stationary and axially symmetric Einstein
 equations. We expect that this method can be used also with other kind of
 boundary conditions.

 \begin{acknowledgments} 
 S. D. thanks Piotr Chru\'sciel for useful discussions. These
 discussions took place at the Institut Mittag-Leffler, during the
 program ``Geometry, Analysis, and General Relativity'', 2008 fall.
 S. D. thanks the organizers of this program for the invitation and
 the hospitality and support of the Institut Mittag-Leffler.

 The authors want to thank Luis Lehner and Steve Liebling for useful discussions
 that took place at FaMAF during their visits in 2008.

 S. D. is supported by CONICET (Argentina).  This work was supported
 in part by grant PIP 6354/05 of CONICET (Argentina), grant 05/B415  Secyt-UNC
 (Argentina) and the Partner Group grant of the Max Planck Institute
 for Gravitational Physics, Albert-Einstein-Institute (Germany).
 \end{acknowledgments} 

 \appendix

 \section{Extreme Kerr solution}\label{sec:extr-kerr-solut}

 The extreme Kerr black hole corresponds to the limit $m=\sqrt{|J|}$
 of the Kerr metric, where $m$ is the total mass and $J$ is the angular
 momentum of the spacetime. Usually, instead of $J$ in the literature
 the parameter $a=J/m$ is used,  the extreme limit correspond to
 $a=\pm m $.

 Using the notation of section \ref{sec:vari-probl-parab}, for the
 extreme Kerr black hole  we have only one end $i_0$ located at the
 origin. 
 The explicit form of the functions
  $(\x_0,\Y_0)$ are given by
 \begin{equation}
   \label{eq:79b}
 \x_0=\log \X_0-2 \log \rho, \quad \Y_0 = \hat \Y_0-
 \frac{2 J^3\cos\theta\sin^4\theta}{|J| \Sigma},  
 \end{equation}
 where
 \begin{align}
   \label{eq:57b}
 \X_0 & =\left(\tilde r^2+|J| +\frac{2|J|^{3/2} \tilde r \sin^2\theta
   }{\Sigma} \right)\sin^2\theta, \\
   \hat \Y_0 & = 2J(\cos^3\theta-3\cos\theta)  \label{eq:57c},
 \end{align}
 and
 \begin{equation}
   \label{eq:58b}
 \tilde r = r+\sqrt{|J|}, \quad 
 \Sigma=\tilde r^2+|J| \cos^2 \theta.
 \end{equation}
 In these equations, $(r,\theta)$ are spherical coordinates in $\Rt$
 related with the cylindrical coordinates $(\rho, z)$ by the standard
 formulas  $r=\sqrt{\rho^2 + z^2}$ and $ \tan \theta =\rho/z$. 

 In this equations,  $J$ is an arbitrary constant. It gives the 
 angular momentum and it is the only free
 parameter in this solution. In agreement with  equation (\ref{eq:27}),
 we have
 \begin{equation}
   \label{eq:16}
   \omega_0(\theta=0)=-4J, \quad \omega_0(\theta=\pi)=4J.
 \end{equation}
 Note that the angular momentum is given by $\hat \omega_0$, the other
 part of $\omega_0$ vanishes at the axis. 

 The singular behavior of  $\sigma_0$ at $i_0$ is given by
 \begin{equation}
   \label{eq:25}
   \sigma_0=-2\log r +O(1).
 \end{equation}

 The sign change $J\to -J$ implies $\x \to \x $ and $\Y
 \to -\Y$.
 The limit $J=0$ correspond to flat spacetime and it is given by
 \begin{equation}
   \label{eq:12}
   \x_0=0, \quad \eta_0=\rho^2, \quad \omega_0=0.
 \end{equation}

 The important property of the functions $(\x_0,\Y_0)$ is that they are
 solutions of equations (\ref{eq:el1})--(\ref{eq:el2}). In the above equations
 the end point $i_0$ is chosen to be at the origin of the coordinate system. We
 have the obvious freedom to translate this point to an arbitrary location. In
 particular, the extreme Kerr solution centered at the point $i_1$ used in
 section \ref{sec:Techniques} is given by
 \begin{equation}
   \label{eq:22}
   \x_1=\x_0(\rho,z-L/2), \quad  \Y_1=\Y_0(\rho,z-L/2).
 \end{equation}


\end{document}